\documentclass[12pt,letterpaper]{article}

\usepackage{fullpage}
\usepackage{enumerate}
\usepackage{amsthm}
\usepackage{amsmath,amssymb}
\usepackage{mathtools}
\usepackage{graphics,graphicx}
\usepackage{subcaption}
\usepackage{booktabs}
\usepackage{multirow}
\usepackage{multicol}
\usepackage{fancyhdr}
\usepackage{textcomp}
\usepackage{setspace}
\usepackage{amsfonts}
\usepackage{upgreek}
\usepackage{dsfont}
\usepackage{natbib}
\usepackage{hyperref}
\usepackage{tikz}
\usepackage{array}
\usepackage{float}
\usepackage{dsfont}
\usepackage{booktabs} 
\usepackage{dirtytalk}
\usepackage[ruled]{algorithm2e} 

\SetAlFnt{\small}
\SetAlCapFnt{\small}
\SetAlCapNameFnt{\small}
\SetAlCapHSkip{0pt}
\IncMargin{-\parindent}

\usepackage{subcaption}

\newtheorem{definition}{Definition}
\newtheorem{assumption}{Assumption}
\newtheorem{proposition}{Proposition}
\newtheorem{lemma}{Lemma}
\newtheorem{theorem}{Theorem}
\newtheorem{remark}{Remark}
\newtheorem{claim}{Claim}
\newtheorem{corollary}{Corollary}

\begin{document}

\onehalfspacing
\title{A Characterization for Optimal Bundling of  Products with Non-Additive Values}

\date{\today}
\author{
  Soheil Ghili\thanks{E-mail: soheil.ghili@yale.edu. Click \href{https://sites.google.com/d/15EgEigNDlM3jOW0yqHBpZE0MHdjZeVGf/p/1F-C3Vr4w1OgoxdNeHXt40LgqE77XaxDP/edit}{here} for the most current version. I thank Dirk Bergemann, Nima Haghpanah, Johannes Horner, Peter Klibanoff, Barry Nalebuff, Larry Samuelson, Kai Hao Yang, Jidong Zhou, and seminar participants for helpful comments. All errors are mine.} \\
	Yale University
}

\maketitle

\begin{abstract}
This paper studies optimal bundling of  products with non-additive values. Under  monotonic preferences and single-peaked profits, I show a monopolist finds pure bundling optimal if and only if the optimal sales volume for the grand bundle is  larger than the optimal sales volume for any smaller bundle.  I then (i)  detail how my analysis relates to ``ratio monotonicity'' results on bundling; and (ii)  describe the implications for non-linear pricing.
\end{abstract}

\section{Introduction}

This paper  studies optimal bundling decisions by a multi-product monopolist, and carries out an analysis with two main features. First,  I allow for non-additive values: a consumer's valuation for a given product can depend on whether s/he has purchased other products as well. Second, I seek to obtain a \textit{full characterization} of when pure bundling (i.e., the act of selling only the package of all available products together as one bundle) is optimal. Under monotonic preferences and single-peaked profits, I prove that optimal bundling admits a simple characterization: Pure bundling is optimal if the optimal sales volume for the grand bundle  (if sold alone) is strictly larger than that for any other bundle. Conversely, if there is at least one bundle whose optimal sales volume (if sold alone) is strictly larger than that of the grand bundle, then pure bundling is sub-optimal.\footnote{Note that this is slightly short of a full characterization given it does not determine the optimal bundling strategy when the optimal sales volume for the grand bundle is larger than those for smaller bundles but not always strictly. This small gap can be closed but with stronger assumptions that I decided not to make. See section \ref{sec: model} for more details.} In simpler terms, bundling is optimal if and only if it helps sell more.

The rest of the paper is organized as follows. Section \ref{sec: lit} reviews the related literature. Section \ref{sec: model} sets up the model and formally presents the assumptions and the main result. Section \ref{sec: proof of main result} provides the proof. Section \ref{sec: discussion} discusses the assumptions and delves into the implications and interpretations of the main result. It also discusses additional results which are provided in the appendix. Section \ref{sec: conclusion} concludes.

\section{Related Literature}\label{sec: lit}
The study of bundling dates at least as far back as \cite{stigler1963united}. Most papers in this literature focus on the case of ``additive values,'' meaning the valuation by each consumer of any given product $i$ is not impacted by whether she also possesses product $i'\neq i$. Pioneering in this area was \cite{adams1976commodity}, pointing out that bundling can be more profitable than unbundling when there is negative correlation among consumers in how they value individual products. Other studies such as \cite{mcafee1989multiproduct,menicucci2015optimality,pavlov2011optimal,schmalensee1984gaussian,fang2006bundle,manelli2007multidimensional,palfrey1983bundling,bergemann2021optimality} further develop results on optimal bundling (or optimal upgrade pricing) under additive values. Many of these studies focus on a setting with two products only. Also, with few exceptions-- e.g., \cite{daskalakis2017strong} who provide necessary and sufficient conditions--most of these studies concentrate on sufficient conditions for bundling. Although most of this literature examines a monopolist seller (which is also the focus of this paper), some studies have analyzed multiple sellers (\cite{mcafee1989multiproduct,zhou2017competitive,zhou2019mixed}).

The literature on non-additive values, to which this paper belongs, is considerably smaller. Part of this literature focuses directly on bundling (e.g., \cite{haghpanah2019pure,armstrong2013more,armstrong2016nonlinear,long1984comments}) whereas some study  price discrimination settings which have implications for bundling (e.g., \cite{anderson2009price,deneckere1996damaged}). This paper complements the literature on non-additive values in that it imposes a different set of assumptions (stronger only than those imposed by \cite{haghpanah2019pure}) and delivers simple-to-interpret but \textit{necessary and sufficient} conditions for bundling based on optimal quantities sold. I also  connect the interpretation of my results to those based on price elasticities (such as \cite{long1984comments,armstrong2013more}) as well as those based on ratio monotonicity (such as \cite{haghpanah2019pure,anderson2009price,deneckere1996damaged,salant1989inducing}). Finally, I describe implications of my bundling results for screening models and nonlinear tariff design (e.g., \cite{mussa1978monopoly,maskin1984}). 

\section{Main Result}\label{sec: model}
\subsection{Setup and Notations}

A monopolist has $n$ products to sell, indexed 1 through $n$. Possible bundles of these products are denoted $b\subseteq\{1,...,n\}$. Set $\mathcal{B}=\{b|b\subseteq\{1,...,n\}\} $ represents the set of all possible bundles.\footnote{My notation, in part, follows \cite{haghpanah2019pure}.} By $\Bar{b}$ denote the grand bundle $\{1,...,n\}$. For any bundle $b$, denote $b^C=\Bar{b}\setminus b$. There is a unit mass of customers whose types are represented by $t\in T\subset \mathbb{R}^m$ where $T$ is compact. As will be shown later, one of the model assumptions will imply that types are one-dimensional (i.e., $m=1$.)\footnote{The reason why I start with general  $m$ and later show $m=1$  is that this exposition clarifies that $m=1$ is a an implication of the rest of the model assumptions, rather than a separate assumption itself.} Probability distribution over types $f(\cdot)>0$ has no atoms.\footnote{The main result should hold without these assumptions on $f$. But I expect the proof to be less clean.} The valuation by type $t$ for bundle $b$ is denoted $v(b,t)$. Assume $v(\emptyset,t)=0$. Also, for all $b$, suppose that $v(b,t)$ is continuous in $t$ except, possibly, for finitely many points. The per-unit cost of production for each product $i$ is $c_i\geq 0$. 

The problem the monopolist solves has two components. First, she makes a bundling decision. She chooses the optimal set $B^*$ of bundles $b$ among subsets $B$ of $\mathcal{B}$ that satisfy $\emptyset\notin B$. Note that there are  $2^{2^n-1}$ possible bundling strategies. Thus, characterizing the conditions under which the monopolist can simply choose $B^*=\{\Bar{b}\}$ should indeed be of value. 

The second decision by the firm is choosing prices $p(\cdot):B\rightarrow \mathbb{R}$ for the bundles offered.\footnote{Note that, in principle, one could model the bundling decision through pricing; because not offering a product would be equivalent to pricing it so high that no customer would purchase it. As such, separating the bundling and pricing decisions in the model is redundant. Nevertheless, I decided to carry out this separation because it makes the notation easier.}  Denote by $\mathcal{P}_B$ the set of all possible such pricing functions.


Once the firm decides on set $B$ and prices $p(\cdot)$, customers decide which bundles to purchase (note that this model only considers deterministic selling procedures.) Each customer $t$'s decision $\beta(t|B,p)\subseteq B$ is determined by:

\begin{equation}\label{eq: customer bundle choice}
    \beta(t|B,p)=\arg\max_{\hat{\beta}\subseteq B} v(\cup_{b\in \hat{\beta}}b,t)-\Sigma_{b\in \hat{\beta}} p(b)
\end{equation}

Throughout, I assume customers break ties in favor of the seller. Also, note that equation \ref{eq: customer bundle choice} implies that customers want at most one unit of each product $i$ and find additional units redundant.

Demand for bundle $b$ is given by the measure of customers $t$ who would choose to purchase bundle $b$:

\begin{equation}\label{eq: demand}
    D(b|B,p)=\int_t \mathds{1}_{b\in \beta(t|B,p)} f(t) dt
\end{equation}

Firm profit under strategy $(B,p)$ is:

\begin{equation}\label{eq: firm profit}
    \pi(B,p)=\int_t \Sigma_{b\in \beta(t|B,p)}\bigg(p(b)-\Sigma_{i\in b}c_i\bigg)  f(t) dt
\end{equation}


The monopolist  chooses $(B^*,p^*)$ to maximize profit:

\begin{equation}\label{eq: firm's problem}
    (B^*,p^*)=\arg\max_{B\in\mathcal{B}, p\in\mathcal{P}_B} \pi(B,p)
\end{equation}

With the setup of the firm problem laid out, I now introduce a few more definitions and notations. For disjoint bundles $b$ and $b'$, denote by $v(b,t|b')$ the valuation by type $t$ for $b$ conditional on possessing $b'$. Formally:

$$v(b,t|b')\equiv v(b\cup b' ,t)-v(b',t)$$

In a similar manner, denote  $\beta(t|B,p,b')=\arg\max_{\hat{\beta}\subseteq (B\setminus\{b'\})} v(\cup_{b\in \hat{\beta}}b,t|b')-\Sigma_{b\in \hat{\beta}} p(b)$. Also denote $D(b|B,p,b')=\int_t 1_{b\in \beta(t|B,p,b')} f(t) dt$. Moreover, denote $$\pi(B,p|b')=\int_t \Sigma_{b\in \beta(t|B,p,b')}\bigg(p(b)-\Sigma_{i\in b}c_i\bigg)  f(t) dt.$$ 

Finally, for any set $\beta$ of bundles, denote $\beta^\cup=\cup_{b\in\beta}b$. Similarly, I write $\beta^\cup(t|B,p,b')$ instead of $\cup_{b\in \beta(t|B,p,b')}b$. I next turn to the assumptions and the main result.

\subsection{Assumptions and Characterization}
The main results of the paper is about how optimal bundling decisions are informed by the comparison among optimal sales volumes for different bundles. I start with some necessary assumptions and definitions.

\begin{assumption}\label{assumption: monotonicity}
\textbf{Monotonicity:} For all $b$, $v(b,t)$ is increasing in $v(\bar{b},t)$, and strictly so whenever $v(b,t)>0$. The same applies to $v(b,t|b^C)$ for all $b$. 
\end{assumption}

\begin{assumption}\label{assumption: quasiconcavity}
\textbf{Quasi-concavity:} For any $b\in \mathcal{B}$, profit functions $\pi(b,p)$ and $\pi(b,p|b^C)$ are strictly quasiconcave in $p(b)$ for all values of $p(b)$ that yield strictly positive demand for $b$.\footnote{This simply means the profit peaks only once as we vary each price.} 
\end{assumption}

\begin{definition}\label{def: optimal quantity}
By $D^*(b)$ denote the ``optimal quantity sold'' of bundle $b$ if no other bundle were offered by the firm. Formally,  $D^*(b)$ is defined as $D(b|\{b\},p_b^*)$ where $p_b^*$ is the optimal price for bundle $b$ when $B=\{b\}$.
\end{definition}

\begin{definition}\label{def: pure bundling}
A given firm strategy $(B,p)$ involves pure bundling if:

$$\forall t: \beta^\cup(t|B,p)  \in \{\emptyset, \Bar{b} \}$$
\end{definition}


 We are now ready to state the main result.


\begin{theorem}\label{theorem: main result}
Under assumptions \ref{assumption: monotonicity} and \ref{assumption: quasiconcavity}, the optimal strategy $(B^*,p^*)$ involves pure bundling if:

    $$D^*(\Bar{b}) > \max_{b\in\mathcal{B}\setminus\{\Bar{b}\}} D^*(b)$$
    
Conversely, the optimal strategy does not involve pure bundling if:

$$D^*(\Bar{b}) < \max_{b\in\mathcal{B}\setminus\{\Bar{b}\}} D^*(b)$$
\end{theorem}

In words, this result says that the firm should pure bundle if and only if it helps ``sell more.''\footnote{Note that this result is slightly short of a full characterization because it does not specify whether pure bundling is optimal when $D^*(\Bar{b}) = \max_{b\in\mathcal{B}\setminus\{\Bar{b}\}} D^*(b)$. One can show that under this last possibility, pure bundling is optimal if instead of assuming profits are strictly quasi-concave in each price, we assume they are strictly concave and differntiable at  peak. Even though this would yield a full characterization, I decided that the ability to speak to the ``measure-zero'' case of $D^*(\Bar{b}) = \max_{b\in\mathcal{B}\setminus\{\Bar{b}\}} D^*(b)$ is too small a return to justify such a restrictive assumption as strict concavity. As such, I maintain the quasi-concavity assumption.}



\section{Proof of Theorem \ref{theorem: main result}}\label{sec: proof of main result}
 
I start by some remarks, definitions, and lemmas.

\begin{lemma}\label{lem: order in t}
There is a mapping $\tau$ from the set $T$ of types $t$ on to the interval $[0,1]$ such that:

\begin{enumerate}
    \item $\forall t,t'\in T: v(\Bar{b},t)>v(\Bar{b},t') \Leftrightarrow \tau(t)>\tau(t')$.
    \item $\tau$ is a sufficient statistic: Once $\tau(t)$ is known, one can fully pin down all $v(b,t)$  without having to know $t$.
    
\end{enumerate}

\end{lemma}

\textbf{Proof of Lemma \ref{lem: order in t}.} Set $\tau(t)\triangleq \frac{v(\Bar{b},t)-\min_{t'\in T}v(\Bar{b},t')}{\max_{t'\in T}v(\Bar{b},t')-\min_{t'\in T}v(\Bar{b},t')}$. By construction, it satisfies (1). To see why it satisfies (2), first note that by monotonicity, for any $t,t'$ such that $v(\Bar{b},t)=v(\Bar{b},t')$, we have $v(b,t)=v(b,t')$ for any other $b$. As a result, $v(\Bar{b},t)$ is sufficient information for determining $v(b,t)$ for all $b$. Next, observe that  one can recover $v(\Bar{b},t)$ from $\tau(t)$: $v(\Bar{b},t)=\min_{t'\in T}v(\Bar{b},t')+\tau(t)\times (\max_{t'\in T}v(\Bar{b},t')-\min_{t'\in T}v(\Bar{b},t'))$. As a result, once one knows $\tau(t)$, one would also know $v(b,t)$ for all $b$. \textbf{Q.E.D.}

Based on this lemma, it is without loss to think of $t$ as $\tau(t)$ and, hence, the set of all possible $t$ as $[0,1]$. Therefore, we can use expressions such as $t\geq t'$. Going forward, I assume $t\in[0,1]$. 

\begin{remark}\label{remark: quasi-concavity}
Suppose functions $f_1(x)$, $f_2(x)$ and $f_1(x)+f_2(x)$ are all strictly quasi-concave over the interval $[a,b]$. Then either (i) $\arg\max f_1\leq \arg\max f_2$ or (ii) $\arg\max f_1\leq \arg\max (f_1+f_2)$ will imply: $$\arg\max f_1\leq \arg\max (f_1+f_2)\leq \arg\max f_2.$$
\end{remark}

The proof is left to the reader.

\begin{definition}\label{def: optimal conditional quantity}
For disjoint bundles $b$ and $b'$, denote  by $D^*(b|b')$ the ``optimal quantity sold'' of bundle $b$ if all customers are already endowed with $b'$ and only $b$ is offered by the firm at optimal price. Formally,  $D^*(b|b')\equiv D(b|\{b\},p_{b|b'}^*,b')$ where $p_{b|b'}^*:\{b\}\rightarrow \mathbb{R}$ is   effectively one real number, and it is chosen among other possible $p$ so that $\pi(\{b\},p|b')$ is maximized.
\end{definition}

Next, I show that the problem of finding the optimal price for a bundle is equivalent to the problem of finding the right type $t^*$ and sell to types $t\geq t^*$.

\begin{definition}\label{definition: optimal pricing equiv optimal type}
Define by $t^*(b|b')$ the largest $t$ such that $1-F(t)\geq D^*(b|b')$. Also, for simplicity, denote $t^*(b|\emptyset)$ by $t^*(b)$.
\end{definition}

\begin{lemma}\label{lem: optimal pricing equiv optimal type}
Consider bundles $b$ and $b^C=\bar{b}\setminus b$. Suppose that all types are endowed with bundle $b^C$, and that the firm is selling only bundle $b$, optimally choosing $p^*_{b|b^C}$. The set of types who will buy the product at this price is the interval $[t^*(b|b^C),1]$.
\end{lemma}

\textbf{Proof of Lemma \ref{lem: optimal pricing equiv optimal type}.} Follows directly from monotonicity. Monotonicity implies that the optimal sales volume $D^*(b|b^C)$ would be purchased by the highest types $t$ with $t$ weakly above some cutoff $\hat{t}$. Definition \ref{definition: optimal pricing equiv optimal type} says that for the demand volume to equal $D^*(b|b^C)$, the cutoff $\hat{t}$ has to equal $t^*(b|b^C)$. \textbf{Q.E.D.}

Lemma \ref{lem: optimal pricing equiv optimal type} is important in that it shows the problem of choosing $p^*_{b|b^C}$ can equivalently be thought of as the problem of choosing $t^*_{b|b^C}$.  This allows us to set up the firm's problem based on $t$. Next definition introduces a necessary notation for this purpose.

\begin{definition}\label{def: profit as function of quantity}
Consider disjoint bundles $b$ and $b'$. Suppose that all types have already been endowed with $b'$, and that the firm is to sell only bundle $b$. By $\pi_b(t|b')$ denote the profit to the firm if it chose a price for bundle $b$ such that all types $t'\geq t$ would purchase bundle $b$:

$$\pi_b(t|b')= (1-F(t))\times \big((v(t,b|b')-\Sigma_{i\in b}c_i\big)$$
$$= \pi(\{b\},v(b,t|b')|b')$$
\end{definition}

\begin{lemma}\label{lem: quasiconcavity in t}
$\pi_b(t|b^C)$ is strictly quasi-concave in $t$.
\end{lemma}

\textbf{Proof of Lemma \ref{lem: quasiconcavity in t}. } Suppose $\pi_b(t|b^C)$ is not quasi-concave in $t$. This means there are $t_1<t_2<t_3$ such that $\pi_b(t_2|b^C)\leq\min(\pi_b(t_1|b^C),\pi_b(t_3|b^C))$. Then construct $p_1,p_2$ and $p_3$ from $t_1, t_2$ and $t_3$ according to the procedure in definition \ref{def: profit as function of quantity}. That is, set $p_i=v(b,t|b^C)$ for each $i$. Monotonicity puts $p_2$ strictly between $p_1$ and $p_3$. Note that for these prices, we have:

$$\pi(\{b\},p_2|b^C)\leq\min(\pi(\{b\},p_1|b^C),\pi(\{b\},p_3|b^C))$$

which violates the quasi-concavity assumption in $p$. \textbf{Q.E.D.}

With the above definitions and lemmas in hand, we are ready to prove the main theorem. I start by the necessity condition (i.e., the condition that $D^*(\Bar{b})\geq D^*(b)$ for all $b$ is necessary for pure bundling to optimal).

\textbf{Proof of necessity.} We want to show that if there is some $b$ such that $D^*(b)>D^*(\Bar{b})$, then pure bundling is sub-optimal. Specifically, I show that offering bundles $b$ and $\Bar{b}$ would be strictly more profitable to the firm compared to offering  $\Bar{b}$ alone. The argument follows.

\begin{lemma}\label{lem: partial bundle conditional vol lower}
$D^*(b)>D^*(\Bar{b})$ implies $D^*(b)>D^*(b^C|b)$.
\end{lemma}

\textbf{Proof of Lemma \ref{lem: partial bundle conditional vol lower}.} Suppose, on the contrary, that $D^*(b)\leq D^*(b^C|b)$. This means $t^*(b)\geq t^*(b^C|b)$. We know:

$$t^*(b)=\arg\max_t \pi_b(t)$$
and
$$t^*(b^C|b)=\arg\max_t \pi_{b^C}(t|b).$$

Also, given definition \ref{def: profit as function of quantity}, it is straightforward to verify that:

$$\pi_{\Bar{b}}(t)\equiv\pi_{b^C}(t|b)+\pi_b(t)$$

By strict quasi-concavity of all profits in $t$ and by remark \ref{remark: quasi-concavity}, it has to be that the argmax of $\pi_{\Bar{b}}(t)$ falls in between the argmax values $t^*(b^C|b)$ and $t^*(b)$. Therefore, we get: $t^*(\Bar{b})\leq t^*(b)$, which implies $D^*(b)\leq D^*(\Bar{b})$, contradicting a premise of the lemma. \textbf{Q.E.D.}

\begin{lemma}\label{lem: proof of sufficiency}
Selling $D^*(b^C|b)$ units of the grand bundle $\bar{b}$ along with $D^*(b)-D^*(b^C|b)$ units of bundle $b$ would be strictly more profitable to the monopolist compared to selling $D^*(\Bar{b})$ units of the grand bundle alone.
\end{lemma}

\textbf{Proof of Lemma \ref{lem: proof of sufficiency}.} Note that given monotonicity and given Lemma \ref{lem: partial bundle conditional vol lower}, selling $D^*(b^C|b)$ units of the grand bundle $\bar{b}$ along with $D^*(b)-D^*(b^C|b)$ units of bundle $b$ would simply mean selling $b$ to types $[t^*(b),t^*(b^C|b))$  and  selling $\bar{b}$ to types $[t^*(b^C|b),1]$. This can be implemented by offering bundles $b$ and $\bar{b}$ and pricing them at $p^*_b$ and $p^*_b+p^*_{b^C|b}$ respectively.\footnote{Recall that I use the notation $p^*_b$ for the optimal price of $b$ when only $b$ is offered. Similarly, $p^*_{b^C|b}$ is the optimal price of $b^C$ when everyone is endowed by $b$ and the monopolist is only selling $b^C$.} This would deliver the following profit:

$$\pi_1=\pi_{b^C}(t^*(b^C|b)|b)+\pi_b(t^*(b))$$

Again, by monotonicity, selling $D^*(\Bar{b})$ units of the grand bundle can be thought of as selling  $\bar{b}$ to types $t^*(\bar{b})$ and above. This would deliver a profit of $\pi_2=\pi_{\bar{b}}(t^*(\bar{b}))$, which can be expanded and written as:

$$\pi_2=\pi_{b^C}(t^*(\bar{b})|b)+\pi_b(t^*(\bar{b}))$$

Note that each term in $\pi_2$ is weakly less than its corresponding term in $\pi_1$ (due to the optimality of the terms in $\pi_1$). Also by the fact that $t^*(b^C|b)>t^*(b)$, then either $t^*(b^C|b)\neq t^*(\bar{b})$ or $t^*(b)\neq t^*(\bar{b})$. Thus, by strict quasi-concavity, at least one of the two inequalities between corresponding terms in $\pi_1$ and $\pi_2$ has to be strict, yielding $\pi_1>\pi_2$. \textbf{Q.E.D.}

Given this lemma, the proof of the if side of the theorem is now complete. \textbf{Q.E.D.}

Note that the proof of the first side did \textit{not} use the constant marginal costs assumption. Next, I turn to the proof of the sufficiency conditions (i.e., that $D^*(\Bar{b})> \max_{b\in\mathcal{B}\setminus\{\Bar{b}\}} D^*(b)$  implies that pure bundling is optimal).

\textbf{Proof of sufficiency.} I start with some lemmas.

\begin{lemma}\label{lem: all products sell}
Under assumptions \ref{assumption: monotonicity} and \ref{assumption: quasiconcavity}, there is a firm optimal strategy $(B^*,p^*)$ such that non-measure-zero set of  customers $t$ we have $\beta^\cup(t|B^*,p^*)=\Bar{b}$.
\end{lemma}

\textbf{Proof of Lemma \ref{lem: all products sell}.} I start by assuming that no optimal strategy $(B^*,p^*)$ involves selling $\Bar{b}$ to a non-measure-zero set of consumers. Then I reach a contradiction by constructing a weakly profitable deviation from an assumed optimal $(B^*,p^*)$ such that the deviation sells $\Bar{b}$ to a non-measure-zero set of consumers.

By $D^*(\bar{b})>D^*(b)$ for all $b\neq\bar{b}$, we get: $D^*(\bar{b})>0$, which in turn yields $t^*(\bar{b})<1$. Next, note that the number of possibilities for $\beta(t|B^*,p^*)$ is finite.  By this finiteness and by piece-wise continuity of value functions, there is some $\tilde{t}\geq t^*(\bar{b})$ such that all consumers with types higher than $\tilde{t}$ have the same purchase behavior.\footnote{Perhaps with the exception of type $t=1$; but that does not matter given its zero measure.} Formally:

$$\forall t,t'\in(\tilde{t},1): \beta^\cup(t|B^*,p^*)=\beta^\cup(t'|B^*,p^*) $$

Denote this commonly purchased bundle $\tilde{b}$. Given that our contrapositive assumption is that no non-measure-zero set of types purchases the grand bundle $\Bar{b}$, it has to be that $\tilde{b}\neq\Bar{b}$. Next, I construct a profitable deviation for the monopolist from $(B^*,p^*)$. First, note that given that currently no consumer purchases $\Bar{b}$ or constructs it from other bundles, it has to be that either $\Bar{b}$ is not part of $B^*$ or it is expensive enough for no customer to prefer to obtain it. Next, assume that the monopolist deviates from $(B^*,p^*)$ by adding $\Bar{b}$ to the set of bundles and pricing it at $p^*(\tilde{b})+\Sigma_{i\in\tilde{b}^C}c_i+\epsilon$ where $\epsilon$ is  chosen so that (i) type $\tilde{t}$ would weakly prefer  $\tilde{b}$ over $\Bar{b}$ but (ii) type 1 would weakly prefer $\Bar{b}$ over $\tilde{b}$. Next, I proceed to show two things. First: finding such an $\epsilon$ is feasible. Second: with that $\epsilon$, the monopolist will see a weak profit increase relative to $(B^*,p^*)$ and sell $\bar{b}$ to a positive-measure set of types.

For type  $\tilde{t}$ to weakly prefer $\tilde{b}$ over $\Bar{b}$, it has to be that 

$$v(\Bar{b},\tilde{t})-p^*(\tilde{b})-\Sigma_{i\in\tilde{b}^C}c_i-\epsilon\leq v(\tilde{b},\tilde{t})-p^*(\tilde{b})$$

$$\Leftrightarrow \epsilon\geq v(\tilde{b}^C,\tilde{t}|\tilde{b})-\Sigma_{i\in\tilde{b}^C}c_i$$

Similarly, for type to 1 weakly to prefer to purchase $\Bar{b}$, one can show that $\epsilon$ must satisfy:

$$\epsilon\leq v(\tilde{b}^C,1|\tilde{b})-\Sigma_{i\in\tilde{b}^C}c_i$$

But by monotonicity, we have $v(\tilde{b}^C,\tilde{t}|\tilde{b})\leq v(\tilde{b}^C,1|\tilde{b})$. Therefore, $\epsilon$ may be chosen within the following interval:

$$[v(\tilde{b}^C,\tilde{t}|\tilde{b})-\Sigma_{i\in\tilde{b}^C}c_i,v(\tilde{b}^C,1|\tilde{b})-\Sigma_{i\in\tilde{b}^C}c_i]$$

If the interval is not a singleton, choose $\epsilon$ in the interior. 

Next, I show that once such $\epsilon$ is chosen, the new bundling and pricing strategy by the firm will weakly increase the profit to the insurer while selling $\Bar{b}$ to a non-measure zero set of consumers. To see the latter, denote by $\tilde{t}'$ the set of types who weakly prefer $\Bar{b}$ over $\tilde{b}$ under this new strategy. By the choice of $\epsilon$ and by monotonicity, we have $\tilde{t}\leq\tilde{t}'<1$. Therefore the new strategy will  change the purchase behavior by types $t\geq \tilde{t}'$ (and only those types.) Next note that this behavior-change is weakly profitable to the monopolist. Prior to this change, the profit to the monopolist from these types was:

$$\pi_1=(1-\tilde{t}')\times(p^*(\tilde{b})-\Sigma_{i\in \tilde{b}}c_i)$$

Under the new strategy (i.e., with the introduction of $\Bar{b}$ at the price of $p^*(\tilde{b})+\Sigma_{i\in\tilde{b}^C}c_i+\epsilon$,) the new profit level from these types is:

$$\pi_2=(1-\tilde{t}')\times\bigg((p^*(\tilde{b})+\Sigma_{i\in\tilde{b}^C}c_i+\epsilon)-\Sigma_{i\in\Bar{b}}c_i\bigg)$$

$$=\pi_1+(1-\tilde{t}')\times \epsilon$$

Thus, it remains to show $\epsilon\geq 0$. To this end, note that by $D^*(\bar{b})>D^*(\tilde{b})$ we have $t^*(\bar{b})\leq t^*(\tilde{b})$. This, by monotonicity, quasi-concavity, and remark \ref{remark: quasi-concavity}, implies $t^*(\tilde{b}^C|\tilde{b})\leq t^*(\bar{b})$ which in turn yields $t^*(\tilde{b}^C|\tilde{b})\leq \tilde{t}$. That is, all types weakly above $\tilde{t}$ would purchase $\tilde{b}^C$ if (i) they were offered it at the optimal price for the monopolist and (ii) they were pre-endowed with $\tilde{b}$. This implies that $\forall t\geq \tilde{t}: v(\tilde{b}^C,t|\tilde{b})\geq p^*_{\tilde{b}^C|\tilde{b}}\geq \Sigma_{i\in\tilde{b}^C}c_i$. But this means $\epsilon=v(\tilde{b}^C,\tilde{t}|\tilde{b})- \Sigma_{i\in\tilde{b}^C}c_i\geq 0$. As a result, we get $\pi_2\geq\pi_1$, which completes the proof of the lemma.\textbf{Q.E.D.}

Next, it is useful to observe that the monotonicity assumption imposes a vertical relationship not only on consumers' preferences, but also on their purchase behaviors.

\begin{lemma}\label{lem: vertical differentiation among customers}
Consider bundling strategy $B$ and pricing strategy $p$. Consider types $t$ and $t'$ such that $\beta^\cup(t|B,p)\neq\beta^\cup(t'|B,p)$. Then the following statements hold:

\begin{enumerate}
    \item If $\beta^\cup(t'|B,p)=\emptyset$, we have $t'<t$.
    \item If $\beta^\cup(t'|B,p)=\Bar{b}$, we have $t'>t$.
\end{enumerate}
\end{lemma}

This lemma says that if there are types who buy $\Bar{b}$, they are the highest types. Also if there are types who buy nothing, they are the lowest types.

\textbf{Proof of Lemma \ref{lem: vertical differentiation among customers}.} I prove the second statement in the lemma. The first statement would be proved in a similar way. Suppose that $\beta^\cup(t|B,p)\neq \beta^\cup(t'|B,p)=\Bar{b}.$ For simplicity, denote $\beta^\cup(t|B,p)=\tilde{b}$. Now suppose, contrary to the statement of the lemma, that $t'\leq t$. Given $\beta(t|B,p)^\cup\neq \beta^\cup(t'|B,p)$, we know $t\neq t'$ which implies $t' < t$. Next, observe the following two inequalities:

First, note that under $(B,p)$, type $t$  prefers purchasing $\beta(t|B,p)$ and forming $\tilde{b}$ over purchasing $\beta(t'|B,p)$ and forming $\Bar{b}$. Formally:

\begin{equation}\label{Lemma: purchase monotonicity eq 1}
    v(\tilde{b},t)-\Sigma_{b\in \beta(t|B,p)}p(b) \geq v(\bar{b},t)-\Sigma_{b\in \beta(t'|B,p)}p(b)
\end{equation}

Similarly, type $t'$  prefers to purchase $\beta(t'|B,p)$ and forming $\bar{b}$ over purchasing $\beta(t|B,p)$ and forming $\tilde{b}$. Formally:

\begin{equation}\label{Lemma: purchase monotonicity eq 2}
    v(\bar{b},t')-\Sigma_{b\in \beta(t'|B,p)}p(b) \geq v(\tilde{b},t')-\Sigma_{b\in \beta(t|B,p)}p(b)
\end{equation}

At least one of the two inequalities above has to be strict (because if both types were indifferent between $\bar{b}$ and $\tilde{b}$, they would break this tie the same way.) Adding these two inequalities together, we get:

$$v(\bar{b},t')+v(\tilde{b},t)> v(\bar{b},t)+v(\tilde{b},t')$$

$$\Leftrightarrow v(\bar{b},t')-v(\tilde{b},t')> v(\bar{b},t)-v(\tilde{b},t)$$

$$v(\tilde{b}^C,t'|\tilde{b})> v(\tilde{b}^C,t|\tilde{b})$$

This latter statement, combined with $t'<t$, contradicts monotonicity. \textbf{Q.E.D.}

In light of lemma \ref{lem: all products sell}, the following two corollaries of lemma \ref{lem: vertical differentiation among customers} are useful.

\begin{corollary}\label{corollary: those who buy b_bar}
Under $(B^*,p^*)$, the set of types to for which $\beta^\cup(t|B^*,p^*)=\Bar{b}$ takes the form of $[t_1,1]$ for some $t_1<1$.
\end{corollary}

\begin{corollary}\label{corollary: those who buy nothing}
Under $(B^*,p^*)$, the set of types to for which $\beta^\cup(t|B^*,p^*)=\emptyset$ takes the form of $[0,t_2)$ for some $t_2<1$.
\end{corollary}

With these lemmas in hand, I next turn to the proof of the sufficiency conditions. The strategy is, again, contrapositive.

Assume on the contrary that we have, at the same time: (i) $\forall b\neq\Bar{b}: D^*(b)< D^*(\Bar{b})$ and (ii) the firm's optimal strategy does not involve pure bundling. This latter statement implies that the set of all distinct bundles chosen by customers under $(B^*,t^*)$ includes members other than $\emptyset$ or $\Bar{b}$. Formally, if we denote $$\beta^*=\{b|\exists t: \beta^\cup(t|B^*,p^*)=b\}$$

then $\beta^{*}\setminus \{\emptyset,\Bar{b}\}\neq \emptyset$. In other words, our contrapositive assumption implies that $t_1$ in corollary \ref{corollary: those who buy b_bar} is strictly larger than $t_2$ in corollary \ref{corollary: those who buy nothing}.

Then, note that by corollary \ref{corollary: those who buy b_bar} and the piece-wise continuity of values in $t$, there is some bundle $b_1\in \beta^*\setminus \{\emptyset,\Bar{b}\}$ such that for $t_1'$ close enough to but smaller than $t_1$, we have:

\begin{equation}\label{eq: t1'}
  \forall t\in[t_1',t_1): \beta^\cup(t|B^*,p^*)=b_1  
\end{equation}

Also, by corollary \ref{corollary: those who buy nothing} and by piece-wise continuity of values in $t$, there is some bundle $b_2\in \beta^*\setminus \{\emptyset,\Bar{b}\}$ (which may or may not be the same as $b_1$) such that for $t_2'$ close enough to but larger than $t_2$, we have:

\begin{equation}\label{eq: t2'}
  \forall t\in[t_2,t_2']: \beta^\cup(t|B^*,p^*)=b_2
\end{equation}

The rest of the proof of the sufficiency conditions of the theorem is organized as follows. I first make a series of claims (without proving them). Next I use the claims to prove the sufficiency conditions of the theorem. Finally, I will return to the proofs of the claims.

\begin{claim}\label{claim: if statement 1}
$t^*(b_1^C|b_1)=t_1$.
\end{claim}

In words, claim \ref{claim: if statement 1} says that the set of customers who purchase the grand bundle $\beta(t|B^*,p^*)=\Bar{b}$ under the firm optimal strategy $(B^*,p^*)$ is the same as those who purchase $b_1^C$ and construct the grand bundle if (i) everyone is endowed with $b_1$ and (ii) the firm offers only $b_1^C$, pricing it  optimally.

\begin{claim}\label{claim: if statement 2}
$t^*(b_2)=t_2$.
\end{claim}

Claim \ref{claim: if statement 2} says that the set of customers who purchase $\emptyset$ under the firm optimal strategy $(B^*,p^*)$ is the same as those who purchase $\emptyset$ if the firm offers only $b_2$ and prices it optimally.

Next, note that the assumption $D^*(\Bar{b})> D^*(b_2)$, combined with monotonicity and claim \ref{claim: if statement 2}, implies $t^*(\Bar{b})\leq t_2$. By $t_1>t_2$, we get $t^*(\Bar{b})<t_1=t^*(b_1^C|b_1)$. Also note that:

$$\forall t: \pi_{\Bar{b}}(t)=\pi_{b_1^C}(t|b_1)+\pi_{b_1}(t)$$

As such, by strict quasi-concavity of profits, by $t^*(\Bar{b})<t^*(b_1^C|b_1)$, and by remark \ref{remark: quasi-concavity}, the peak of $\pi_{\Bar{b}}(t)$ should happen  in between those of $\pi_{b_1^C}(t|b_1)$ and $\pi_{b_1}(t)$. Therefore, we should have: $t^*(b_1)\leq t^*(\Bar{b})\leq t^*(b_1^C|b_1)$. But $t^*(b_1)\leq t^*(\Bar{b})$  implies: $$D^*(b_1)\geq D^*(\Bar{b})$$

which is a contradiction. Therefore, the sufficiency part of the theorem is true provided that claims \ref{claim: if statement 1} and \ref{claim: if statement 2} are true. I now turn to the proofs of these claims.

\textbf{Proof of Claim \ref{claim: if statement 1}.} Suppose on the contrary that $t^*(b_1^C|b_1)\neq t_1$. In that case, it can be shown that the firm can strictly improve its profit upon the optimal strategy $(B^*,p^*)$. The proof of this claim constructs such improvement. To this end, the following remark is useful to state.

\begin{remark}\label{rem: singleton choices}
Construct the bundling strategy $(\hat{B},\hat{p})$ from $(B^*,p^*)$ in the following way:
\begin{itemize}
    \item $\hat{B}=\{\beta^\cup(t|B^*,p^*) \forall t\}$
    \item For each $t$, or in other words for each $\hat{b}=\beta^\cup(t|B^*,p^*)\in \hat{B}$, set $\hat{p}(\hat{b})=\Sigma_{b\in\hat{b}}p^*(b)$.
\end{itemize}

For such $(\hat{B},\hat{p})$, we have:

\begin{enumerate}
    \item $\forall t: \beta(t|\hat{B},\hat{p})=\beta(t|B^*,p^*) $
    \item $\pi(\hat{B},\hat{p})=\pi(B^*,p^*)$
\end{enumerate}

\end{remark}

This remark simply states that there is an optimal strategy by the seller under which each buyer type only purchases a single bundle instead of combining different bundles to construct her/his desired one. I skip the proof of this remark. Also, in order to save on notation, I assume from now on that it is $(B^*,p^*)$ itself that has the feature of every $\beta(t|B^*,p^*)$ being a singleton. I now return to the proof of claim \ref{claim: if statement 1} and construct a strict improvement upon the profit of $(B^*,p^*)$.

I do so by slightly adjusting the price of $\Bar{b}$. That is, I show that there is a pricing strategy  $p$ with $p(b)=p^*(b)$ for all $b\neq \Bar{b}$, but with $p(\Bar{b})\neq p^*(\Bar{b})$, such that $\pi(B^*,p)>\pi(B^*,p^*)$.

To see why this is the case, construct bundling strategy $B'$ in the following way:

\begin{equation}
    B'=\{b_1,\Bar{b}\}
\end{equation}

Also construct pricing strategy $p'$ by fixing $p'(b_1)=\min_t v(b_1)$ and setting $p'(\Bar{b})=p'(b_1)+\rho$ where $\rho$ is a parameter that we will vary.

More specifically, I show that  as long as  $\rho \in[p^*(\Bar{b})-p^*(b_1)-\epsilon,p^*(\Bar{b})-p^*(b_1)+\epsilon]$ for a small enough $\epsilon$, then $\pi(B^*,p)$ and $\pi(B',p')$ move \textit{in parallel} if we set $p(\Bar{b})=p^*(b_1)+\rho$ and $p'(\Bar{b})=p'(b_1)+\rho$, and move $\rho$ (that is, as we change $\rho$, the difference $\pi(B^*,p)-\pi(B',p')$ remains constant). The range parameter $\epsilon$ should be chosen so that for any pricing strategy $p$ constructed with a $\rho$ in this interval we have: $D(\Bar{b}|B^*,p)<1-F(t_1')$ where $t_1'$ was constructed in equation \ref{eq: t1'}. In other words, $\epsilon$ should be small enough (or, alternatively, $p(\Bar{b})$ should be close enough to $p^*(\Bar{b})$ ) so that as we change $\rho$, the only types who are affected are those around $t_1$; and, hence, the only purchase decisions that are affected are choices between $b_1$ and $\Bar{b}$. 

With this setup, note that if we set $\epsilon=0$, then demand for grand bundle $\Bar{b}$ under both strategies will be equal to demand for the grand bundle under the optimal strategy:

$$\epsilon=0\Rightarrow D(\Bar{b}|B^*,p)=D(\Bar{b}|B',p')=D(\Bar{b}|B^*,p^*)$$

Next, note that for small $\epsilon\neq 0$, we will still have $D(\Bar{b}|B^*,p)=D(\Bar{b}|B',p')$ because such changes in $\rho$ will lead the exact same set of types to switch their purchase decisions between $\Bar{b}$ and $b_1$, under both strategies $(B^*,p)$ and $(B',p')$. This leads to the exact same revenue change between the two strategies as a result of the change in $\epsilon$ (or, equivalently, in $\rho$). Also the change in total costs are the same given the constant-marginal-costs assumption. As a result, $\pi(B^*,p)$ and $\pi(B',p')$ change in the same way as a result of small changes in $\rho$.\footnote{One can show we do not need the constant marginal cost assumption if $n=2$. or if the monotonicity condition is strengthened.}

Now note that if claim \ref{claim: if statement 1} does not hold, then $\pi(B',p')$ under $\epsilon=0$ is not the global maximum for $\pi(B',p')$. By strict quasi-concavity, it is not a local maximum either. As a result, there is a small change in $\rho$ that would strictly increase $\pi(B',p')$. Given that $\pi(B',p')$ and $\pi(B^*,p)$ move in parallel if we slightly change $\rho$, then $\pi(B^*,p)$ should also strictly increase relative to $\pi(B^*,p^*)$, a contradiction.\textbf{Q.E.D.}

\textbf{Proof of Claim \ref{claim: if statement 2}.} The proof of this claim is  similar to that of the previous claim. We start by assuming, on the contrary, that $t^*(b_2)\neq t_2$ and reach a contradiction. Construct $(B',p')$ by assuming $B'=\{b_2\}$, which makes $p'$ just one number (for the price of $b_2$). Similar to the previous claim, one can show that for prices $\rho$ for $b_2$ sufficiently close to $p^*(b_2)$ the two profit functions $\pi(B^*,p)$ and $\pi(B',p')$ move in parallel as we move $\rho$. Again, similarly to the previous claim, this implies that $(B^*,p^*)$ can be improved upon if $t_2\neq t^*(b_2)$. \textbf{Q.E.D.}

The completion of the proofs for claims \ref{claim: if statement 1} and \ref{claim: if statement 2} finishes the proof of the sufficiency side of the theorem, and hence the theorem itself. \textbf{Q.E.D.}

\section{Discussion}\label{sec: discussion}


\subsection{Discussion of the assumptions}\label{subsec: discussion of assumptions}

\textbf{Monotonicity:} The most restrictive assumption in this model is monotonicity, which, as Lemma \ref{lem: order in t} shows, makes the type space uni-dimensional. It is worth noting that monotonicity is quite common in the literature. A prevalent version of it in the screening literature is the single crossing condition imposed by \cite{maskin1984}.\footnote{Single crossing in \cite{maskin1984} is in fact stronger than my monotonicity condition. In order for monotonicity to be as strong as the single crossing condition in \cite{maskin1984}, it has to be that $v(\Bar{b},t')\geq(>)v(\Bar{b},t) \Rightarrow \forall b\cap b'=\emptyset: v(b,t'|b')\geq(>)v(b,t|b')$.} Within the bundling literature most of the papers that focus on products with non-additive values focus on versions of the product-line pricing problem--which can be thought of as a special case of the bundling problem--and each impose a form of monotonicity (e.g., \cite{anderson2009price,deneckere1996damaged,long1984comments}). This usually comes in the form of assumed increasing difference of values in the (unidimensional) type and the product quality level. Also the seminal paper by \cite{mussa1978monopoly} on product line pricing assumes the valuation by each type of each quality level is proportional to both type and quality, which implies monotonicity.\footnote{It is worth noting that unlike other papers on product line pricing such as \cite{anderson2009price}, the \cite{mussa1978monopoly} study does \textit{not} focus on whether and when bundling is optimal. \cite{mussa1978monopoly} make a series of assumptions that imply pure bundling  (i.e., offering only the highest quality version) is always sub-optimal. One can indeed show that an appropriate translation of the \cite{mussa1978monopoly} problem into the setting of this paper will satisfy the optimal sales volume conditions for sub-optimality of pure bundling. See appendix for more details.} To my knowledge, the only studies of bundling of products with non-additive values that do not impose a version of monotonicity are \cite{haghpanah2019pure} and \cite{armstrong2013more}. That said, the ratio-monotonicity conditions in \cite{haghpanah2019pure} --at least in one direction-- have a similar implication to my monotonicity condition.\footnote{To be clear, there are many papers in the bundling literature that do not impose a version of monotonicity as a model assumption or as part of the conditions in their theorems. To the best of my knowledge, however, \textit{all such papers study environments with additive values,} meaning they impose $\forall b,t: v(b,t)=\Sigma_{i\in b}v(\{i\},t)$.}

Finally, note that the monotonicity assumption \textit{does not} force consumers to rank the products the same way. That is, it does not rule out $v(b,t)>v(b',t)$ co-existing with $v(b,t')<v(b',t')$. It, rather, rules out $v(b,t)>v(b,t')$ co-existing with $v(b',t)<v(b',t')$.

\textbf{Quasi-concavity:} Quasi-concavity simply requires that each relevant profit function be single-peaked. This assumption has been made in the literature before (e.g., see assumption 5 in \cite{maskin1984}). Without this assumption, one can still prove a version of Theorem \ref{theorem: main result}; but that version would be weaker and less straightforward to state. Finally, it is worth specifying what this assumption would look like if expressed based on the model primitives rather than profit functions. This assumption would require that $\frac{\partial \log(v(b,t)-\Sigma_{i\in b}c_i)}{\partial t}-\frac{f(t)}{1-F(t)}$ cross zero only once from above for all $b\neq\emptyset$.

\textbf{Other notes:} Most papers on optimal bundling of products with non-additive values assume there are only two products whereas this paper examines arbitrarily many. Additionally, one direction of the results in Theorem \ref{theorem: main result} does not use the assumption that marginal costs are constant. Furthermore, to my knowledge, with the exception of this paper and \cite{haghpanah2019pure} which do not make assumptions on substitution patterns, other papers on bundling with non-additive values impose relatively strong  complementarity assumptions.\footnote{These assumptions are mostly implicit in the form of product-line pricing which may be thought of as a ``base product'' plus complementary add-ons. As an exception, \cite{armstrong2013more} allows for complementarity or substitution separately, but does not allow the same products to be complementary for some types and substitutes for others.}

\subsection{Interpretation of the Result and Relation to Literature}\label{subsec: ratio monotonicity}
In the literature on bundling of products with non-additive values, a commonly mentioned condition for (sub-)optimality of pure bundling is ratio monotonicity.  According to ratio-monotonicity, pure bundling is optimal if $\frac{v(b,t)}{v(\bar{b},t)}$ is \textit{everywhere} weakly increasing in $v(\bar{b},t)$ for all $b$. Also pure-bundling is sub-optimal if this fraction is \textit{everywhere} strictly increasing for at least one $b$. Versions of it have been mentioned in \cite{anderson2009price,salant1989inducing,deneckere1996damaged}, and, most generally, \cite{haghpanah2019pure}.\footnote{\cite{haghpanah2019pure} prove their results under weaker underlying assumptions, use a stochastic version of ratio-monotonicity (which is weaker than the deterministic version and allows for multi-dimensional types,) and are to my knowledge the only paper that states both sides of the condition.} Theorem \ref{theorem: main result} shows that, at the cost of having to make assumptions, \ref{assumption: monotonicity} and \ref{assumption: quasiconcavity}, optimality of pure bundling could be linked to local--as opposed to global--ratio monotonicity.

\begin{proposition}\label{prop: primitives}
Suppose that assumptions \ref{assumption: monotonicity} and \ref{assumption: quasiconcavity} hold and that production costs are zero. Also suppose each $\pi(b,p)$ is differentiable in $p$ and that the derivative is only zero at the peak. Then for any two bundles $b$ and $b'$ we have $D^*(b')> D^*(b)$ if and only if for the unique $\tilde{t}$ with $\frac{\partial \log(v(b',\tilde{t}))}{\partial t}=\frac{f(\tilde{t})}{1-F(\tilde{t})}$, we have: $\frac{\partial \log(v(b,\tilde{t}))}{\partial t}> 
\frac{f(\tilde{t})}{1-F(\tilde{t})}$.
\end{proposition}

The proof follows directly from the assumptions. To see why this is related to local ratio monotonicity, note that according to Proposition \ref{prop: primitives}, for $D^*(\Bar{b})>D^*(b)$ to hold, we need  the $\tilde{t}$ that satisfies $\frac{\partial \log(v(\Bar{b},\tilde{t}))}{\partial t}=\frac{f(\tilde{t})}{1-F(\tilde{t})}$ to also satisfy $\frac{\partial \log(v(b,\tilde{t}))}{\partial t}>\frac{f(\tilde{t})}{1-F(\tilde{t})}$. In other words: $\frac{\partial \log(v(\Bar{b},\tilde{t}))}{\partial t}<\frac{\partial \log(v(b,\tilde{t}))}{\partial t}$. This means that for $t$ slightly larger than $\tilde{t}$, we get: $\frac{v(\Bar{b},t)}{v(\Bar{b},\tilde{t})}<\frac{v(b,t)}{v(b,\tilde{t})}$. Rearranging, we get $\frac{v(b,\tilde{t})}{v(\Bar{b},\tilde{t})}<\frac{v(b,t)}{v(\Bar{b},t)}$. This is exactly ratio monotonicity.\footnote{Note that this also to some extent resembles elasticity-comparison results such as those in \cite{armstrong2013more} and \cite{long1984comments}.} 

Note that, according to this result, in order to check whether pure bundling is optimal, one does \textit{not} need to calculate $D^*(b)$ for all $b$. \textit{It would suffice to calculate $D^*(\bar{b})$ and then check ratio-monotonicity at that local point.}

\textbf{Parametric Examples:} One can construct simple examples in which optimal sales volumes/local ratio monotonicity can pin down optimal bundling strategy whereas global ratio monotonicity cannot. Suppose $n=2$, and $t$ is uniformly distributed between 0 and 1. Assume the firm can produce these products at no cost. By $b$ denote the bundle $\{1\}$. For simplicity, assume the complementary bundle $b^c=\{2\}$ is not valued by any type: $\forall t: v(\{2\},t)=0$. A common example of this is when $\{2\}$ is an ``add on,'' which is not of value by itself but can add value once the ``base product'' is present (e.g., additional memory for a smart phone). Suppose $\forall t: v(b,t)=t+k_1$ where $k_1$ is a fixed real number. Finally, assume  $\forall t: v(\Bar{b},t)=t+k_1+t^{k_2}$ where $k_2$ is a positive real number. That is, each type $t$'s valuation of the add-on on top of the original product is $t^{k_2}$. One can verify that this setup satisfies monotonicity and quasi-concavity.

Now fix $k_1=0.2$ and allow $k_2$ to vary. One can verify that for $k_2\leq\frac{2}{3}$,  pure bundling is optimal while for $k_2>\frac{2}{3}$ it is optimal to mixed-bundle. Also, once $k_2$ passes $\frac{2}{3}$, the optimal sales volume for the grand bundle passes $0.6$ (which is the optimal sales volume for partial bundle $b$) from above. Additionally, $\frac{\partial \log\big(v(\bar{b},t)\big)}{\partial t}-\frac{f(t)}{1-F(t)}$ evaluated at $t=t^*(b)=0.4$ passes 0 from below once $k_2$ passes $\frac{2}{3}$. Finally, one can verify that for an interval around $k_2=\frac{2}{3}$, (say $k_2\in[0.4,0.8]$) the ratio $\frac{v(b,t)}{v(\Bar{b}),t}$ first decreases and then increases as $v(\Bar{b})$ grows. To sum up, the optimal bundling decision is informed by optimal sales volumes/local ratio monotonicity but not by global ratio monotonicity. The appendix provides a visualization for this parametric class of problems.

I finish this section by noting that the relation to ratio-monotonicity implies that the intuition provided by \cite{haghpanah2019pure} is relevant to my framework as well: the \textit{correlation between WTP and perceived complementarity/substitution levels} has implications for bundling decisions. Pure bundling is optimal if higher WTP customers see a lower degree of complementarity (or higher degree of substitution) across products, compared to lower WTP customers.

\subsection{Implications for Non-linear Pricing}\label{subsec: nonlinear pricing}
The traditional screening/non-linear pricing problem (a la \cite{maskin1984} or \cite{mussa1978monopoly}) can be cast as a form of bundling problem in which each product $i$ represents the $i$-th quality/quantity unit. The appendix shows that a version of the sales volumes result in Theorem \ref{theorem: main result} can be used to \textit{fully characterize} what the optimal tariff should look like in a non-linear pricing problem (observe that this goes beyond the result in Theorem \ref{theorem: main result} which, instead of fully characterizing the bundling strategy, only specifies when pure bundling is optimal.) Leaving the details to the appendix, here I only discuss one important implication of it through an example: 

\textbf{Example.} Suppose types $t$ are uniformly distributed within $[0,1]$. The monopolist sells a single product with continuous quality levels $q\in[0,2]$. The monopolist optimally prices each quality level at $p^*(q)$ where $p^*(0)$ is fixed at 0. Production costs are zero. Valuations are given by $v(q,t)=q\sqrt[q]{t}$,\footnote{This function simply captures the idea that the higher $q$, the more concave the value is across types $t$. For instance, $v(0,t)\equiv0$, $v(\frac{1}{2},t)\equiv \frac{t^2}{2}$, $v(1,t)\equiv t$, and $v(2,t)\equiv2\sqrt{t}$.} and each type $t$ decides which quality level to purchase. In this setting, one can show that it is optimal to charge a flat fee of $\frac{2}{\sqrt{3}}=1.15$ for any strictly positive level of quality.  Under this tariff, types $t<\frac{1}{3}$ purchase nothing while the rest of the types purchase the highest quality version of the product $q=1$. To see why this is true, one can apply our main result and check local ratio monotonicity at $t=\frac{1}{3}$: we have $\frac{\partial \log\big(v(q,t)\big)}{\partial t}=\frac{f(t)}{1-F(t)}$ when evaluated at $t=\frac{1}{3}$ and $q=2$. But for $t=\frac{1}{3}$ and any $q<2$, we have $\frac{\partial \log\big(v(q,t)\big)}{\partial t}=\frac{3}{q}>\frac{3}{2}=\frac{f(t)}{1-F(t)}$.\footnote{Of course for our main result, which was developed in the bundling domain, to apply in this non-linear pricing domain, additional propositions are needed. The appendix provides those.}

This example shows a case where the optimal non-linear tariff leads to a ``jump'' across types in the quality purchased (from $q=0$ to $q=1$ at $t=\frac{1}{3}$.) The literature tends to assume this away. \cite{mussa1978monopoly} state that \textit{``[t]he economic rationale for the conclusion that jumps in $q(\theta)$ are not optimal is that the monopolist would not be making full use of his power to discriminate among different types of buyers.''}\footnote{They denote type by $\theta$ instead of $t$ in this paper; and they denote optimal quality purchased by type $\theta$ under optimal contract by $q(\theta)$.}  By applying bundling results to non-linear pricing, this paper offers a different perspective: Changing the tariff from one that induces a continuous range of purchased quantities/qualities  in an interval $[\underline{q},\Bar{q}]$ to one that induces a jump from $\underline{q}$ to $\Bar{q}$ may indeed be optimal, as long as this discontinuity leads sufficiently many consumers to ``upgrade to $\Bar{q}$'' as opposed to ``downgrade to $\underline{q}$.'' In other words, inducing a jump in purchase behavior in tariff design may be optimal for similar economic reasons to those that make pure bundling of products  optimal.

\section{Conclusion}\label{sec: conclusion}

This paper studied when pure bundling is optimal for a monopolist who sells products with non-additive values (i.e., with no restriction on complementarity/substitution patterns.) Under monotonicity and quasi-concavity assumptions, I showed that pure bundling is optimal if and only if the grand bundle, once sold on its own and optimally priced, would ``sell more'' than any smaller bundle. The appendix provides additional results on the relation to non-linear tariff design, and the implications of the model for the case of additive values.

This paper provides the first if-and-only-if characterization for optimality of pure bundling under non-additive values; but at the expense of a monotonicity assumption that makes the type space single-dimensional. A full characterization result under non-additive values and multi-dimensional types would, in my view, be the most important way to extend the results in this paper.

\bibliographystyle{ACM-Reference-Format}
\bibliography{Main}


\begin{thebibliography}{23}


\ifx \showCODEN    \undefined \def \showCODEN     #1{\unskip}     \fi
\ifx \showDOI      \undefined \def \showDOI       #1{#1}\fi
\ifx \showISBNx    \undefined \def \showISBNx     #1{\unskip}     \fi
\ifx \showISBNxiii \undefined \def \showISBNxiii  #1{\unskip}     \fi
\ifx \showISSN     \undefined \def \showISSN      #1{\unskip}     \fi
\ifx \showLCCN     \undefined \def \showLCCN      #1{\unskip}     \fi
\ifx \shownote     \undefined \def \shownote      #1{#1}          \fi
\ifx \showarticletitle \undefined \def \showarticletitle #1{#1}   \fi
\ifx \showURL      \undefined \def \showURL       {\relax}        \fi
\providecommand\bibfield[2]{#2}
\providecommand\bibinfo[2]{#2}
\providecommand\natexlab[1]{#1}
\providecommand\showeprint[2][]{arXiv:#2}

\bibitem[\protect\citeauthoryear{Adams and Yellen}{Adams and Yellen}{1976}]%
        {adams1976commodity}
\bibfield{author}{\bibinfo{person}{William~James Adams} {and}
  \bibinfo{person}{Janet~L Yellen}.} \bibinfo{year}{1976}\natexlab{}.
\newblock \showarticletitle{Commodity bundling and the burden of monopoly}.
\newblock \bibinfo{journal}{\emph{The quarterly journal of economics}}
  (\bibinfo{year}{1976}), \bibinfo{pages}{475--498}.
\newblock


\bibitem[\protect\citeauthoryear{Anderson and Dana~Jr}{Anderson and
  Dana~Jr}{2009}]%
        {anderson2009price}
\bibfield{author}{\bibinfo{person}{Eric~T Anderson} {and}
  \bibinfo{person}{James~D Dana~Jr}.} \bibinfo{year}{2009}\natexlab{}.
\newblock \showarticletitle{When is price discrimination profitable?}
\newblock \bibinfo{journal}{\emph{Management Science}} \bibinfo{volume}{55},
  \bibinfo{number}{6} (\bibinfo{year}{2009}), \bibinfo{pages}{980--989}.
\newblock


\bibitem[\protect\citeauthoryear{Armstrong}{Armstrong}{2013}]%
        {armstrong2013more}
\bibfield{author}{\bibinfo{person}{Mark Armstrong}.}
  \bibinfo{year}{2013}\natexlab{}.
\newblock \showarticletitle{A more general theory of commodity bundling}.
\newblock \bibinfo{journal}{\emph{Journal of Economic Theory}}
  \bibinfo{volume}{148}, \bibinfo{number}{2} (\bibinfo{year}{2013}),
  \bibinfo{pages}{448--472}.
\newblock


\bibitem[\protect\citeauthoryear{Armstrong}{Armstrong}{2016}]%
        {armstrong2016nonlinear}
\bibfield{author}{\bibinfo{person}{Mark Armstrong}.}
  \bibinfo{year}{2016}\natexlab{}.
\newblock \showarticletitle{Nonlinear pricing}.
\newblock \bibinfo{journal}{\emph{Annual Review of Economics}}
  \bibinfo{volume}{8} (\bibinfo{year}{2016}), \bibinfo{pages}{583--614}.
\newblock


\bibitem[\protect\citeauthoryear{Bergemann, Bonatti, Haupt, and
  Smolin}{Bergemann et~al\mbox{.}}{2021}]%
        {bergemann2021optimality}
\bibfield{author}{\bibinfo{person}{Dirk Bergemann}, \bibinfo{person}{Alessandro
  Bonatti}, \bibinfo{person}{Andreas Haupt}, {and} \bibinfo{person}{Alex
  Smolin}.} \bibinfo{year}{2021}\natexlab{}.
\newblock \showarticletitle{The optimality of upgrade pricing}.
\newblock  (\bibinfo{year}{2021}).
\newblock


\bibitem[\protect\citeauthoryear{Daskalakis, Deckelbaum, and Tzamos}{Daskalakis
  et~al\mbox{.}}{2017}]%
        {daskalakis2017strong}
\bibfield{author}{\bibinfo{person}{Constantinos Daskalakis},
  \bibinfo{person}{Alan Deckelbaum}, {and} \bibinfo{person}{Christos Tzamos}.}
  \bibinfo{year}{2017}\natexlab{}.
\newblock \showarticletitle{Strong Duality for a Multiple-Good Monopolist}.
\newblock \bibinfo{journal}{\emph{Econometrica}} \bibinfo{volume}{85},
  \bibinfo{number}{3} (\bibinfo{year}{2017}), \bibinfo{pages}{735--767}.
\newblock


\bibitem[\protect\citeauthoryear{Deneckere and Preston~McAfee}{Deneckere and
  Preston~McAfee}{1996}]%
        {deneckere1996damaged}
\bibfield{author}{\bibinfo{person}{Raymond~J Deneckere} {and}
  \bibinfo{person}{R Preston~McAfee}.} \bibinfo{year}{1996}\natexlab{}.
\newblock \showarticletitle{Damaged goods}.
\newblock \bibinfo{journal}{\emph{Journal of Economics \& Management Strategy}}
  \bibinfo{volume}{5}, \bibinfo{number}{2} (\bibinfo{year}{1996}),
  \bibinfo{pages}{149--174}.
\newblock


\bibitem[\protect\citeauthoryear{Fang and Norman}{Fang and Norman}{2006}]%
        {fang2006bundle}
\bibfield{author}{\bibinfo{person}{Hanming Fang} {and} \bibinfo{person}{Peter
  Norman}.} \bibinfo{year}{2006}\natexlab{}.
\newblock \showarticletitle{To bundle or not to bundle}.
\newblock \bibinfo{journal}{\emph{The RAND Journal of Economics}}
  \bibinfo{volume}{37}, \bibinfo{number}{4} (\bibinfo{year}{2006}),
  \bibinfo{pages}{946--963}.
\newblock


\bibitem[\protect\citeauthoryear{Haghpanah and Hartline}{Haghpanah and
  Hartline}{2021}]%
        {haghpanah2019pure}
\bibfield{author}{\bibinfo{person}{Nima Haghpanah} {and} \bibinfo{person}{Jason
  Hartline}.} \bibinfo{year}{2021}\natexlab{}.
\newblock \showarticletitle{When is pure bundling optimal?}
\newblock \bibinfo{journal}{\emph{The Review of Economic Studies}}
  \bibinfo{volume}{88}, \bibinfo{number}{3} (\bibinfo{year}{2021}),
  \bibinfo{pages}{1127--1156}.
\newblock


\bibitem[\protect\citeauthoryear{Long}{Long}{1984}]%
        {long1984comments}
\bibfield{author}{\bibinfo{person}{John~B Long}.}
  \bibinfo{year}{1984}\natexlab{}.
\newblock \showarticletitle{Comments on" Gaussian Demand and Commodity
  Bundling"}.
\newblock \bibinfo{journal}{\emph{The Journal of Business}}
  \bibinfo{volume}{57}, \bibinfo{number}{1} (\bibinfo{year}{1984}),
  \bibinfo{pages}{S235--S246}.
\newblock


\bibitem[\protect\citeauthoryear{Manelli and Vincent}{Manelli and
  Vincent}{2007}]%
        {manelli2007multidimensional}
\bibfield{author}{\bibinfo{person}{Alejandro~M Manelli} {and}
  \bibinfo{person}{Daniel~R Vincent}.} \bibinfo{year}{2007}\natexlab{}.
\newblock \showarticletitle{Multidimensional mechanism design: Revenue
  maximization and the multiple-good monopoly}.
\newblock \bibinfo{journal}{\emph{Journal of Economic theory}}
  \bibinfo{volume}{137}, \bibinfo{number}{1} (\bibinfo{year}{2007}),
  \bibinfo{pages}{153--185}.
\newblock


\bibitem[\protect\citeauthoryear{Maskin and Riley}{Maskin and Riley}{1984}]%
        {maskin1984}
\bibfield{author}{\bibinfo{person}{Eric Maskin} {and} \bibinfo{person}{John
  Riley}.} \bibinfo{year}{1984}\natexlab{}.
\newblock \showarticletitle{Monopoly with incomplete information}.
\newblock \bibinfo{journal}{\emph{RAND Journal of Economics}}
  \bibinfo{volume}{15}, \bibinfo{number}{2} (\bibinfo{year}{1984}),
  \bibinfo{pages}{171--196}.
\newblock


\bibitem[\protect\citeauthoryear{McAfee, McMillan, and Whinston}{McAfee
  et~al\mbox{.}}{1989}]%
        {mcafee1989multiproduct}
\bibfield{author}{\bibinfo{person}{R~Preston McAfee}, \bibinfo{person}{John
  McMillan}, {and} \bibinfo{person}{Michael~D Whinston}.}
  \bibinfo{year}{1989}\natexlab{}.
\newblock \showarticletitle{Multiproduct monopoly, commodity bundling, and
  correlation of values}.
\newblock \bibinfo{journal}{\emph{The Quarterly Journal of Economics}}
  \bibinfo{volume}{104}, \bibinfo{number}{2} (\bibinfo{year}{1989}),
  \bibinfo{pages}{371--383}.
\newblock


\bibitem[\protect\citeauthoryear{Menicucci, Hurkens, and Jeon}{Menicucci
  et~al\mbox{.}}{2015}]%
        {menicucci2015optimality}
\bibfield{author}{\bibinfo{person}{Domenico Menicucci}, \bibinfo{person}{Sjaak
  Hurkens}, {and} \bibinfo{person}{Doh-Shin Jeon}.}
  \bibinfo{year}{2015}\natexlab{}.
\newblock \showarticletitle{On the optimality of pure bundling for a
  monopolist}.
\newblock \bibinfo{journal}{\emph{Journal of Mathematical Economics}}
  \bibinfo{volume}{60} (\bibinfo{year}{2015}), \bibinfo{pages}{33--42}.
\newblock


\bibitem[\protect\citeauthoryear{Mussa and Rosen}{Mussa and Rosen}{1978}]%
        {mussa1978monopoly}
\bibfield{author}{\bibinfo{person}{Michael Mussa} {and}
  \bibinfo{person}{Sherwin Rosen}.} \bibinfo{year}{1978}\natexlab{}.
\newblock \showarticletitle{Monopoly and product quality}.
\newblock \bibinfo{journal}{\emph{Journal of Economic theory}}
  \bibinfo{volume}{18}, \bibinfo{number}{2} (\bibinfo{year}{1978}),
  \bibinfo{pages}{301--317}.
\newblock


\bibitem[\protect\citeauthoryear{Palfrey}{Palfrey}{1983}]%
        {palfrey1983bundling}
\bibfield{author}{\bibinfo{person}{Thomas~R Palfrey}.}
  \bibinfo{year}{1983}\natexlab{}.
\newblock \showarticletitle{Bundling decisions by a multiproduct monopolist
  with incomplete information}.
\newblock \bibinfo{journal}{\emph{Econometrica: Journal of the Econometric
  Society}} (\bibinfo{year}{1983}), \bibinfo{pages}{463--483}.
\newblock


\bibitem[\protect\citeauthoryear{Pavlov}{Pavlov}{2011}]%
        {pavlov2011optimal}
\bibfield{author}{\bibinfo{person}{Gregory Pavlov}.}
  \bibinfo{year}{2011}\natexlab{}.
\newblock \showarticletitle{Optimal mechanism for selling two goods}.
\newblock \bibinfo{journal}{\emph{The BE Journal of Theoretical Economics}}
  \bibinfo{volume}{11}, \bibinfo{number}{1} (\bibinfo{year}{2011}).
\newblock


\bibitem[\protect\citeauthoryear{Salant}{Salant}{1989}]%
        {salant1989inducing}
\bibfield{author}{\bibinfo{person}{Stephen~W Salant}.}
  \bibinfo{year}{1989}\natexlab{}.
\newblock \showarticletitle{When is inducing self-selection suboptimal for a
  monopolist?}
\newblock \bibinfo{journal}{\emph{The Quarterly Journal of Economics}}
  \bibinfo{volume}{104}, \bibinfo{number}{2} (\bibinfo{year}{1989}),
  \bibinfo{pages}{391--397}.
\newblock


\bibitem[\protect\citeauthoryear{Schmalensee}{Schmalensee}{1984}]%
        {schmalensee1984gaussian}
\bibfield{author}{\bibinfo{person}{Richard Schmalensee}.}
  \bibinfo{year}{1984}\natexlab{}.
\newblock \showarticletitle{Gaussian demand and commodity bundling}.
\newblock \bibinfo{journal}{\emph{Journal of business}} (\bibinfo{year}{1984}),
  \bibinfo{pages}{S211--S230}.
\newblock


\bibitem[\protect\citeauthoryear{Stigler}{Stigler}{1963}]%
        {stigler1963united}
\bibfield{author}{\bibinfo{person}{George~J Stigler}.}
  \bibinfo{year}{1963}\natexlab{}.
\newblock \showarticletitle{United States v. Loew's Inc.: A note on
  block-booking}.
\newblock \bibinfo{journal}{\emph{The Supreme Court Review}}
  \bibinfo{volume}{1963} (\bibinfo{year}{1963}), \bibinfo{pages}{152--157}.
\newblock


\bibitem[\protect\citeauthoryear{Wilson}{Wilson}{1993}]%
        {wilson1993nonlinear}
\bibfield{author}{\bibinfo{person}{Robert~B Wilson}.}
  \bibinfo{year}{1993}\natexlab{}.
\newblock \bibinfo{booktitle}{\emph{Nonlinear pricing}}.
\newblock \bibinfo{publisher}{Oxford University Press on Demand}.
\newblock


\bibitem[\protect\citeauthoryear{Zhou}{Zhou}{2017}]%
        {zhou2017competitive}
\bibfield{author}{\bibinfo{person}{Jidong Zhou}.}
  \bibinfo{year}{2017}\natexlab{}.
\newblock \showarticletitle{Competitive bundling}.
\newblock \bibinfo{journal}{\emph{Econometrica}} \bibinfo{volume}{85},
  \bibinfo{number}{1} (\bibinfo{year}{2017}), \bibinfo{pages}{145--172}.
\newblock


\bibitem[\protect\citeauthoryear{Zhou}{Zhou}{2019}]%
        {zhou2019mixed}
\bibfield{author}{\bibinfo{person}{Jidong Zhou}.}
  \bibinfo{year}{2019}\natexlab{}.
\newblock \showarticletitle{Mixed Bundling in Oligopoly Markets}.
\newblock  (\bibinfo{year}{2019}).
\newblock


\end{thebibliography}

\appendix

\section*{Online Appendix}

The online appendix provides several additions to the paper. Appendix \ref{apx: visualization} provides visualizations for the examples in the paper. Appendix \ref{apx: Nonlinear Pricing} formally discusses the implications of the main results of the paper for non-linear pricing.  Finally, Appendix \ref{apx: additive values} studies the implications of our results (as well as those of the ratio-monotonicity results from the literature) for environments where values are additive.

\section{Visualizations of Examples from the Main Text}\label{apx: visualization}

\begin{figure}
    \centering
    \begin{subfigure}[t]{0.45\textwidth}
        \centering
        \includegraphics[width=\linewidth]{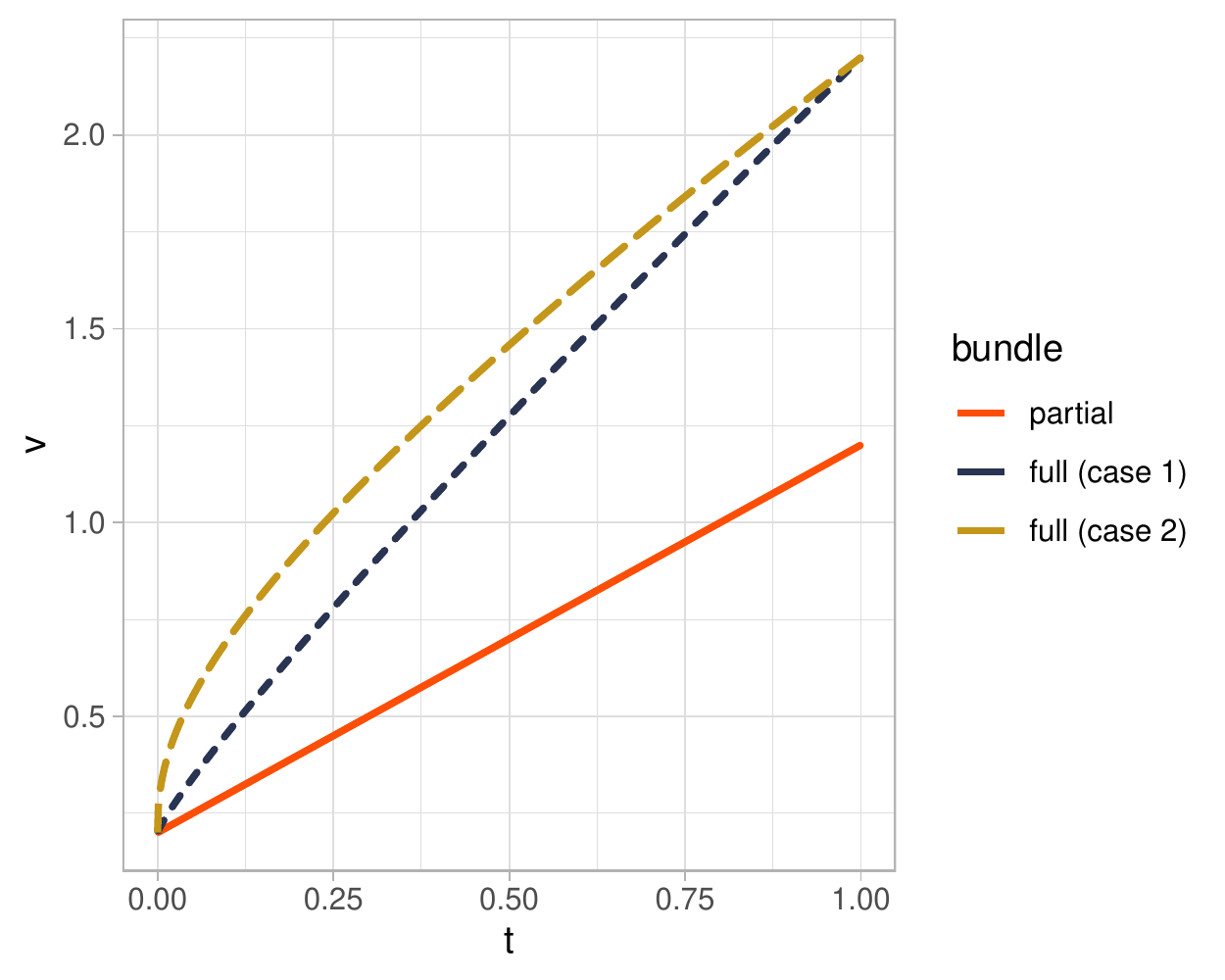} 
        \caption{$k_1=0.2$, Case 1: $k_2=0.4$, Case 2: $k_2=0.8$} \label{subfig:Example 1}
    \end{subfigure}
    \hfill
    \begin{subfigure}[t]{0.45\textwidth}
        \centering
        \includegraphics[width=\linewidth]{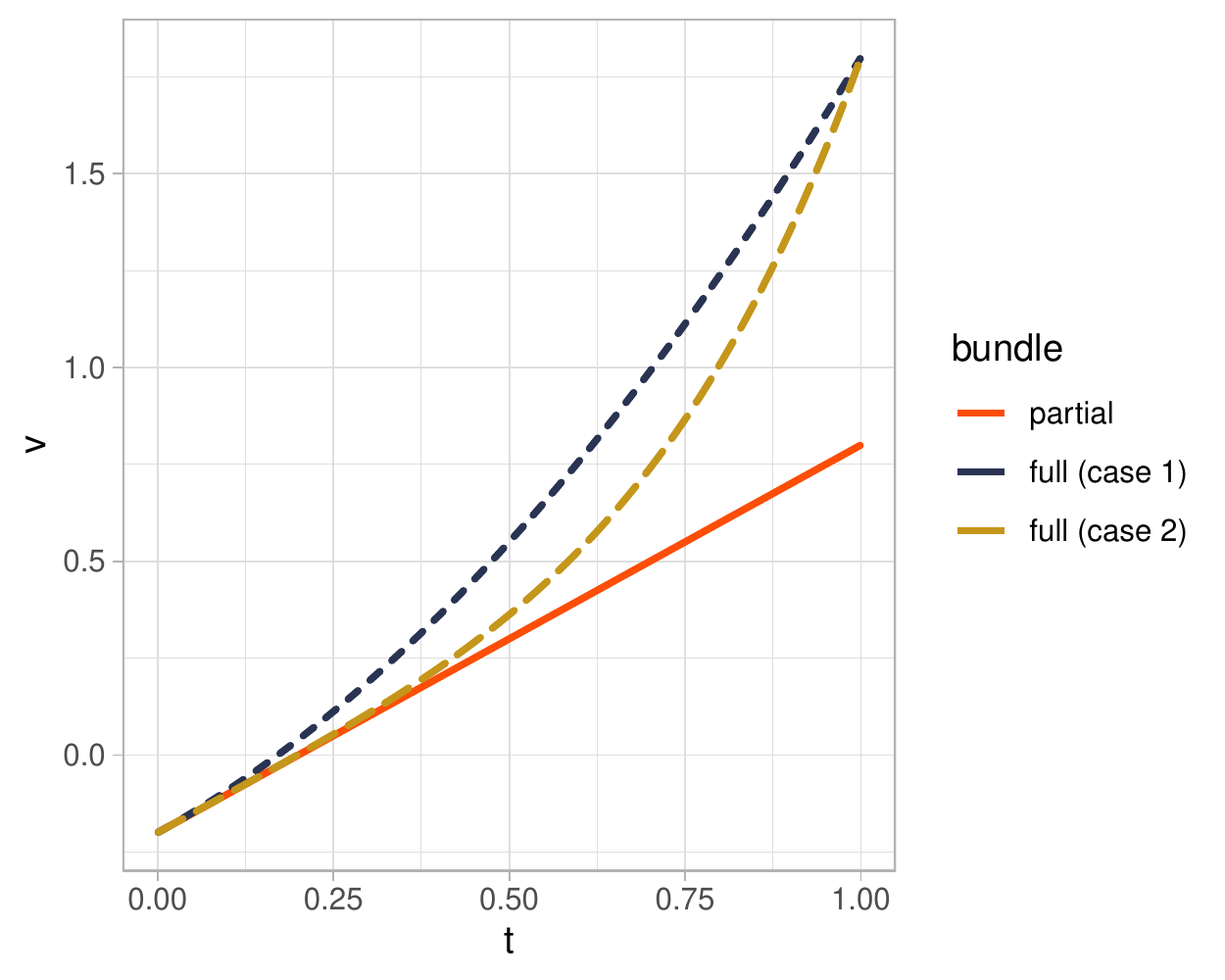} 
        \caption{$k_1=-0.2$, Case 1: $k_2=2$, Case 2: $k_2=1.2$} \label{subfig:Example 2}
    \end{subfigure}
    \caption{Family of examples in which $v(b,t)=t+k_1$ and $v(\bar{b},t)=t+k_1+t^{k_2}$.  }\label{fig: value functions examples}
\end{figure}

The two panels of Figure \ref{fig: value functions examples} visualize the example from Section \ref{subsec: ratio monotonicity} of the paper. As a reminder, the example was about a parametric class of value functions in which $v(b,t)=t+k_1$ and $v(\bar{b},t)=t+k_1+t^{k_2}$. Each panel of the figure has a fixed value for $k_1$ and examines a range of values for $k_2$. For both cases of both panels, the ratio $\frac{v(b,t)}{v(\bar{b},t)}$ is non-monotonic in $v(\bar{b},t)$, making ratio monotonicity results less useful in determining the optimal bundling strategy. However, optimal sales volume results are in line with the optimal bundling decisions. In panel (a), pure bundling is sub-optimal in case 1 and optimal in case 2. In panel (b), pure bundling is optimal in case 1 and sub-optimal in case 2.

\begin{figure}
    \centering
    \includegraphics[scale=.6]{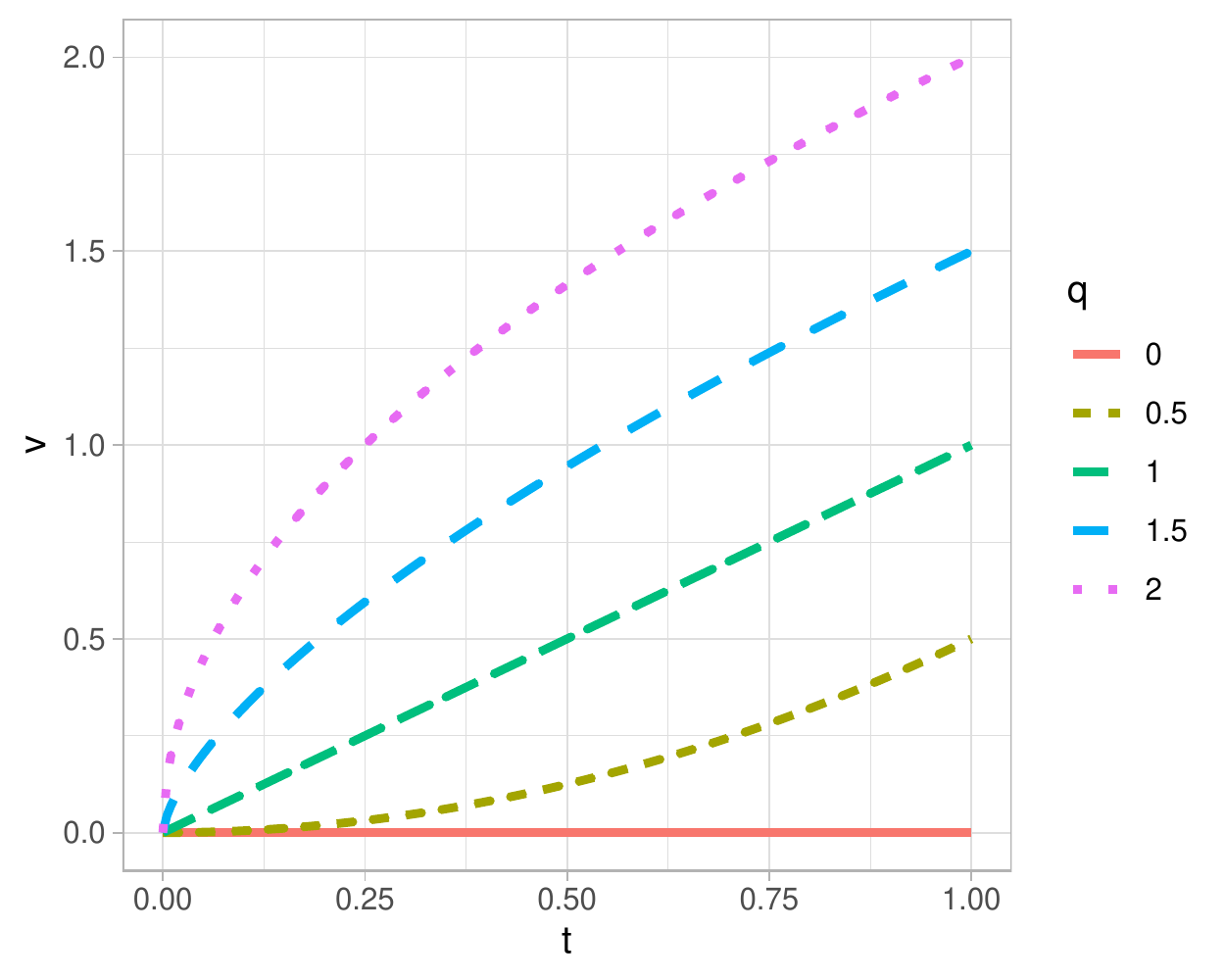}
    \caption{Value function $v(q,t)=q\sqrt[q]{t}$ as a function of $t$ for different values of $q$}
    \label{fig: nonlinear pricing example}
\end{figure}

Also, Figure \ref{fig: nonlinear pricing example} visualizes the value function $v(q,t)=q\sqrt[q]{t}$ from Section \ref{subsec: nonlinear pricing} on the implications of our results for nonlinear tariff design. It plots $v(q,t)$ as a function of $t$ for different values of $q$. As can be seen and also algebraically verified, this function is increasing in both arguments. The key feature of this value function is that it becomes more concave in $t$ as $q$ increases (which implies the optimal sales volume increases in $q$.) Other value functions that share this feature should also yield a similar result, yielding a jump in consumer purchase behavior under the optimal tariff.

\section{Implications for Non-Linear Pricing}\label{apx: Nonlinear Pricing}

The main focus of this paper is the problem of optimal bundling. Nevertheless, the main result also has implications for the non-linear pricing problem studied, among others, by \cite{maskin1984} and \cite{mussa1978monopoly}. Translated to the context of non-linear pricing, the analysis in this paper will take the following form: A monopolist selling a single product faces the problem of determining the optimal price schedule $T^*(q)$ for different quantities $q\in\{0,1,...,n\}$. The monopolist's objective is to maximize total profit $\pi(T^*)$. Valuations are denoted $v(q,t)$ with $v(0,t)=0$ for all $t$. Conditional on a prior endowment of $q'$, we denote the valuations by $v(q,t|q')$. Similarly, we use notation $\pi_q(p)$ to denote the firm's profit when it sells only $q$-size batches of the product, pricing each batch at $p\in\mathbb{R}$. Also $\pi_q(p|q')$ denotes the same thing under the condition that all consumers have been pre-endowed with $q'\leq n-q$ unites of the product. $D^*(q)$, and $D^*(q|q')$ are defined in the expected way.  

Proposition \ref{prop: maskin riley} shows that in the above setting, one can fully characterize the optimal nonlinear tariff.

\begin{proposition}\label{prop: maskin riley}
Consider the non-linear pricing setup described above. Assume the following: (i) $v(q,t)$ is increasing in both arguments (and strictly so whenever positive), and $\forall t'>t, q>0$ we have $v(q,t')-v(q-1,t')>v(q,t)-v(q-1,t)$. (ii) For all $q+q'\leq n$, the profit function $\pi_q(p|q')$ is strictly quasi-concave in $p$ over the range of $p$ that generates strictly positive demand. (iii) $v(q,t)$ is continuous in $t$ except possibly for finitely many points. (iv) For any $q'<n$, the set $\arg\max_{q\in\{1,...,n-q'\}}D^*(q|q')$ is a singleton denoted $q^*(q')$.\footnote{That is, $q^*({q'})$ is the batch-size that, if sold and optimally priced in a market where all consumers have already been endowed with ${q'}$ unites, would generate the highest demand. Similar to the main result, if instead of quasi-concavity we have concavity, one need not impose the assumption that the argmax is unique.} Then the optimal price schedule will involve $m\leq n$ distinct quantities $(q_1^*,..., q_m^*)$ such that $q_1^*=q^*(0)$ and $\forall i\in\{2,...,m\}: q_i^*=q_{i-1}^*+q^*(q_{i-1}^*)$. Precisely:

$$\forall q\in\{q_{i-1}^*+1,...,q_i^*\}: T^*(q)=p^*(q_i^*-q_{i-1}^*|q_{i-1}^*)$$
\end{proposition}

This proposition fully characterizes the optimal tariff $T^*$. It starts by stating that the smallest quantity that any consumer can buy is $q_1^*=q^*(0)\equiv \arg\max_{q\in\{1,...,n\}}D^*(q)$, i.e., the batch size that would \textit{sell the most if sold alone and priced optimally}. In other words, at least $q_1^*$ units of the products are always bundled together. The price of this $q^*_1$-bundle is the optimal price that the monopolist would set if it were to only sell this bundle. From this point on, the proposition takes a recursive structure and states that the second distinct quantity sold of the product is $q_2^*=q_1^*+\arg\max_{q\in\{1,...,n-q_1^*\}}D^*(q|q_1^*)$ and so on. Though Proposition \ref{prop: maskin riley} in some ways resembles the demand profile approach of \cite{wilson1993nonlinear}, it has fundamental differences. Before turning to the proof of this results, I make a few notes.

First, as mentioned shortly before, Proposition \ref{prop: maskin riley} takes a step beyond the main result in Theorem \ref{theorem: main result}: it fully characterizes what the optimal tariff looks like as opposed to only characterizing the conditions under which the optimal tariff will involve a flat fee for selling all $n$ units of the product.\footnote{That condition would be $n=\arg\max_{q\in\{1,...,n\}}D^*(q)$, mirroring the condition in Theorem \ref{theorem: main result}.} This property of Proposition \ref{prop: maskin riley} should naturally raise the question that whether Theorem \ref{theorem: main result} can also be strengthened to fully characterize the optimal bundling strategy as opposed to only characterizing when pure bundling is optimal. Unfortunately the answer turns out to be no; a bundling strategy constructed in a similar way to how the optimal tariff in Proposition \ref{prop: maskin riley} is constructed may or may not be optimal.\footnote{Counterexamples are available upon request. Also, it is worth noting that the Maskin-Riley type monotonicity condition used in Proposition \ref{prop: maskin riley} is stronger than that used in Theorem \ref{theorem: main result}. Nevertheless, this is \textit{not} the reason why the optimal bundling strategy cannot be fully characterized.}

Second, there is a difference between Proposition \ref{prop: maskin riley} and the usual form in which the nonlinear pricing problem is studied in the literature. The origin of Proposition \ref{prop: maskin riley} being in bundling makes the domain of quantities $q$ by construction discrete and bounded, whereas the literature (e.g., \cite{maskin1984}) examines a continuous and possibly unbounded environment. That said, I do not see this feature of Proposition \ref{prop: maskin riley} as too restrictive, given that one an always examine the limit case as $n\rightarrow\infty$.

Fourth, $q$ in this proposition need not be interpreted as quantity. One can also think of $q$ as quality in a similar fashion to \cite{mussa1978monopoly}. In that case, the result holds if the cost function is non-linear in quality, as long as it is still linear in the quantity of the consumers that purchase each quality level. 

Next, I turn to proving the proposition.

\textbf{Proof of Proposition \ref{prop: maskin riley}.}
First, let us introduce some notations. In a similar spirit to the consumer choice notation $\beta(t|B,p)$ in the bundling problem, define the following consumer choice function:

\begin{equation}\label{eq: consumer choice quantity}
    \beta(t|T)\equiv \arg\max_{q} v(t,q)-T(q)
\end{equation}

I use the same notation $\beta$ as I did in the formulation of the bundling problem in order for the parallels between the two settings to be clearer. However, there are some differences. Most notably, the output of the $\beta$ function here is just an integer number, not a bundle. Suppose consumers break ties in favor of higher quantities. I now proceed to state the following lemma.

\begin{lemma}\label{lem: q monotonicity}
For any price schedule $T$ and any two types $t,t'$ with $t'>t$ we have: $\beta(t'|T)\geq \beta(t|T)$.
\end{lemma}

This lemma, whose proof is rather straightforward and is left to the reader, simply says that the consumption quantity is weakly monotonic in type.

Also, in this proof, I will assume that the optimal schedule $T^*(q)$ is weakly monotonic in $q$. This assumption is without loss, given that one can show that any non-monotonic $T$ can be modified in a way that (i) makes it weakly monotonic and (ii) delivers the same amount of profit to the monopolist. Given this weak monotonicity, $T^*$ has to take the following form for a strictly increasing sequence $q_0=0,q_1,...,q_m=n$ and a weakly increasing sequence $T_1,...,T_m$:

\begin{equation}\label{eq: monotone form of TStar}
    \forall q\in\{q_{i-1}+1,...,q_i\}: T^*(q)=T_i
\end{equation}

The following lemma will be useful for the proof:

\begin{lemma}\label{lem: q1 optimal quantity}
$T_1=p^*(q_1|0)$. That is, the lowest type $t$ purchasing $q_1$ under the optimal contract will satisfy $1-F(t)=D^*(q_1)$.
\end{lemma}

\textbf{Proof of Lemma \ref{lem: q1 optimal quantity}.} Suppose this lemma's claim is not true: $T_1\neq p^*(q_1|0)$. Then consider a scenario in which \textit{only} $q_1$ is being sold by the seller at the price of $T_1$. Given that $T_1$ is not the optimal price, there is a small but nonzero $\epsilon$ such that if the seller prices $q_1$ at $T_1+\epsilon$ instead of $T_1$, the seller will strictly improve its profit when selling only $q_1$. 

Next, I move from the scenario of selling only a batch of $q_1$ units back to the full tariff design problem. I use the above deviation to construct a similar deviation from the full schedule $T^*$ and show that the seller can strictly improve its profit. Construct price schedule $T\equiv T^*+\epsilon$ for any positive $q$. I now claim  that $T$ is strictly more profitable to the seller than is $T^*$.

To see why this claim is true, note that for any $q\in\{q_2,...,q_m\}$ and any type $t$ such that $\beta(t|T^*)=q$, we have $\beta(t|T)=q$. This is because (i) for those types it is the IC constraint (and not the IR) that is binding; and (ii) the construction of $T$ from $T^*$ preserves all the IC constraints. Therefore a move from $T^*$ to $T$ will lead to the exact same revenue change that a move from $T_1$ to $T_1+\epsilon$ does in the sceniario of selling only $q_1$: the exact same new types are added to (or removed from) the set that purchases $q_1$, and the same change (i.e., $\epsilon$) has been made to the amount made off of each type that buys. In addition to the change in the revenue, a move from $T^*$ to $T$ also leads to the exact same change in total costs as would a change from $T_1$ to $T_1+\epsilon$. This is because in both cases, the only change made in the production is the number of $q_1$-size batches (or, in the \cite{mussa1978monopoly} interpretation, the number of $q_1$-quality products). Given that we have assumed the cost function to be linear in this change, the changes in total cost is the same between the two scenarios. 

As a result, a move from $T^*$ to $T$ will lead to the exact same change in the total profit as would a move from $T_1$ to $T_1+\epsilon$. Therefore, a change from $T^*$ to the new schedule $T$ strictly profitable, finishing the proof of the lemma. \textbf{Q.E.D.}

Next, I introduce a lemma which will be the building block of the proof of this proposition.

\begin{lemma}\label{lem: q building block}
In the presentation of $T^*$ in equation \ref{eq: monotone form of TStar}, it has to be that $q_1=q_1^*$ where $q_1^*=q^*(0)$ as defined in the statement of Proposition \ref{prop: maskin riley}.
\end{lemma}

As a reminder, $q^*(0)=\arg\max_{q=1,...,n}D^*(q)$. Thus, the lemma simply says that the smallest quantity that the optimal schedule $T^*$ offers to consumers is the quantity that, if sold alone, would sell the highest volume.

\textbf{Proof of Lemma \ref{lem: q building block}.} Suppose $q_1\neq q^*_1$. Then one can construct a deviation from $T^*$ that would strictly improve the seller's profit. This will suffice to finish the proof of the lemma.

First, for convenience, assume that even though $q_1\neq q^*_1$, there is some $k>1$ such that $q_k=q^*_1$. In other words, even though $q^*_1$ is not the smallest package on the schedule, it is nonetheless somewhere on the schedule. Later I will show this assumption is not necessary. But for now, it will make the steps of the proof more straightforward.

For each $i\in\{1,...,m\}$ denote by $t_i$ the lowest type that buys $q_i$ units under $T^*$. That is: $t_i=\min\{t:\,\beta(t|T^*)=q_i\}$.  From Lemma \ref{lem: q monotonicity} we know larger types buy weakly more units of the product. That is: $t_1<t_2<...<t_m$.

From Lemma \ref{lem: q1 optimal quantity} we know that $1-F(t_1)=D^*(q_1)$. Given $t_{k-1}\geq t_1$, we get $1-F(t_{k-1})\leq D^*(q_1)$. But we also know, by assumption, that $D^*(q^*_1)>D^*(q_1)$. Therefore:

\begin{equation}\label{eq: q lemma 1}
    1-F(t_{k-1})<D^*(q^*_1)\equiv D^*(q_k)
\end{equation}

In addition, again by the definition of $q_k=q_1^*$, we know that $D^*(q_{k-1})<D^*(q_k)$. This inequality, along with a similar quasi-concavity argument to that used in the proof of the main result yields:

\begin{equation}\label{eq: q lemma 2}
    D^*(q_k)\leq D^*(q_k-q_{k-1}|q_{k-1})
\end{equation}

Together, inequalities \ref{eq: q lemma 1} and \ref{eq: q lemma 2} yield:

\begin{equation}\label{eq: q lemma 3}
    1-F(t_{k-1})< D^*(q_k-q_{k-1}|q_{k-1})
\end{equation}

Obviously, by $t_k>t_{k-1}$, we also know:

\begin{equation}\label{eq: q lemma 4}
    1-F(t_{k})< D^*(q_k-q_{k-1}|q_{k-1})
\end{equation}

Note that equation \ref{eq: q lemma 4} implies that if all consumers have already been endowed with $q_{k-1}$ units of the product and the monopolist is selling only batches of size $q_k-q_{k-1}$ and pricing them so that types $t_k$ and above purchase, then the monopolist, by quasi-concavity, will strictly profit from a small price reduction $\rho$.

Next, I move from the scenario of selling only $q_k-q_{k-1}$ packages under a pre-endowment of $q_{k-1}$ to the main scenario of designing the full schedule $T^*$. I argue that the monopolist will enjoy the same profit increase as the one described in the previous paragraph if it modifies $T^*$ by reducing all $T_j$ for $j\geq k$ by $\rho$. The argument is similar to that in the proof of Lemma \ref{lem: q1 optimal quantity}. This modification does not alter the behavior of any type $t$ that purchases $q_{k+1}$ or more units due to the fact that it preserves all of the binding IC constraints for those types. As a result, the impact of this change on the firm profit is (i) a cut in margin by $\rho$ across all consumers, combined by the change in the behavior of those who used to purchase less than $q_k$ units under $T^*$ but will now switch to $q_k$.\footnote{None of these types would switch to buying more than $q_k$ units, due to the single-crossing condition.} Note that if the price reduction $\rho$ is small enough,  these types will only consist of those who under $T^*$ purchase $q_{k-1}$. From equation \ref{eq: q lemma 3}, we know there is a non-zero mass of such types. Therefore, the effect of this price change parallels that of the price change described in the previous paragraph, making it strictly profitable. 

But the above argument contradicts the optimality of $T^*$. Therefore, the contrapositive assumption must have been incorrect. That is: it has to be that $q_1=q^*_1$. 

With the above argument, the proof is complete for the case where there is some $k>1$ with $q_k=q^*_1$. That is, when $q_1^*$ is ``on the price schedule.'' Thus, it remains to show that the proof also works when $\nexists k: q_k=q^*_1$. Suppose this is the case, and take $k$ to be the smallest index with $q_k\geq q^*_1$. Like before, denote by $t_i$ the smallest type that purchases $q_i$ under $T^*$

I now construct schedule $T^{**}$ in the following way:

\begin{itemize}
    \item For any $q\leq q_{k-1}$ or $q>q^*_1$, set $T^{**}(q)=T^*(q)$
    \item For all $q_{k-1}<q\leq q^*_1$, set $T^{**}(q)=T^*(q_k)+v(q_1^*,t_k)-v(q_k,t_k)$
\end{itemize}

Next, I take two steps. First, I show that $T^{**}$ delivers the same profit to the monopolist as does $T^*$. Then, I will construct a deviation from $T^{**}$ that yields a strict profit improvement.

As for the first step, note that by construction, $T^{**}(q^*_1)$ is designed to make the type $t_k$ consumer indifferent between purchasing $q_k$ units and $q^*_1$ units. But we know, by construction of $t_k$, that this type is also indifferent between buying $q_k$ units and buying $q_{k-1}$ units. This makes this type indifferent among all three quantities $q_{k-1}<q^*_1<q_k$ under the tariff $T^{**}$. But by the monotonicity condition (i.e., single crossing,) any type $t<t_k$ will strictly prefer $q_{k-1}$ to $q^*_1$ and any type $t>t_k$ will strictly prefer $q_k$ over $q^*_1$. In other words, this ``addition of $q^*_1$ to the schedule'' will not change any consumer's purchase behavior: $\forall t: \beta(t|T^*)=\beta(t|T^{**})$. Thus the two tariffs deliver the same profit to the monopolist.

But now the structure of $T^{**}$ allows us to construct the same profit enhancing modification that we applied to $T^*$ when we assumed it did have $q^*_1$ on the schedule. All of the steps are the same. This finishes the proof of the lemma.
\textbf{Q.E.D.}

The rest of the proof is straightforward and involves recursive use of Lemma \ref{lem: q building block}. First note that by quasi-concavity, and by $\forall q\in\{1,...,n-q^*_1\}: D^*(q+q_1^*)<D^*(q_1^*)$, we have:

$$\forall q\in\{1,...,n-q^*_1\}: D^*(q|q_1^*)<D^*(q_1^*)$$

Thus, all of the optimal strategies will sell only to types $t_1$ and above. This means that in order to construct ``the rest of the optimal schedule'' conditional on having set $T^*(q)$ for all $q\leq q^*_1$ equal to $p^*(q|0)$, one can just focus attention on designing the optimal schedule for selling $n-q^*_1$ units when all consumers have been endowed with $q^*_1$ units already. This means we are facing a version of the same problem. It is straightforward to check that all of the conditions of the main problem are satisfied for this ``sub-problem.'' Thus, one can  apply Lemma \ref{lem: q building block} again and find that $q_2$ in the optimal schedule should be equal to $q^*_2$ which was defined as $q^*_1+q^*(q^*_1)$. Repeating this procedure will fully characterize the optimal tariff and the outcome matches what the statement of the proposition predicted. \textbf{Q.E.D.}

Note that Proposition \ref{prop: maskin riley} can now be used to characterize the optimal contract in the non-linear pricing example given in the main text (i.e., the one in Section \ref{subsec: nonlinear pricing} with $v(q,t)\equiv q\sqrt[q]{t}$.) That example could be cast as a limit case of Proposition \ref{prop: maskin riley} as $n\rightarrow\infty$. It can be seen that the quality level that would generate the highest sales volume would be $q=2$. To check this, it would be sufficient to verify that ratio monotonicity holds at $t=\frac{1}{3}$ which is the lowest type that would purchase if only quality $q=2$ is offered and optimally priced.

\section{A Brief Analysis of Environments with Additive Values}\label{apx: additive values}

In this section, I explore the implications of Theorem \ref{theorem: main result} for environments with additive values (i.e., environments in which $\forall b,t: v(b,t)=\Sigma_{i\in b}v(\{i\},t)$), and compare that to implications of ratio monotonicity conditions that are used in the literature.

Unfortunately, both ratio monotonicity conditions and conditions of Theorem \ref{theorem: main result} are of limited use when reduced to environments with additive values. This point is especially pronounced about the former. The results below, further clarify this matter.

\begin{proposition}\label{prop: additive ratio monotonicity}
Suppose values are additive. Do not impose assumptions \ref{assumption: monotonicity}-\ref{assumption: quasiconcavity} but instead assume ratio monotonicity: $$\forall t,t': v(\bar{b},t')\geq v(\bar{b},t)\Rightarrow \frac{v(b,t')}{v(\bar{b},t')}\geq(\leq) \frac{v(b,t)}{v(\bar{b},t)}$$

Under both of these weak forms of ratio-monotonicity (i.e., $\geq$ or $\leq$,) all  value functions are proportional. That is, for all $b,t,t'$ we have: $\frac{v(b,t)}{v(\bar{b},t)}=\frac{v(b,t')}{v(\bar{b},t')}$. As a corollary, strict ratio monotonicity under additive values is infeasible.
\end{proposition}

\textbf{Proof.} Define $\alpha(i,t)=\frac{v(\{i\},t)}{v(\bar{b},t)}$. By ratio monotonicity, we have $\forall i: \alpha(i,t')\geq\alpha(i,t)$. But we also know by additivity of values that $\Sigma_{i=1,...n}\alpha(i,t')=\Sigma_{i=1,...n}\alpha(i,t)=1$. As a result, it has to be that $\forall i: \alpha(i,t')=\alpha(i,t)$. In other words:

$$\forall i: \frac{v(\{i\},t')}{v(\bar{b},t')}=\frac{v(\{i\},t)}{v(\bar{b},t)}$$

By additivity of values, we can say the same for all $b$:

$$\forall b: \frac{v(b,t')}{v(\bar{b},t')}=\frac{v(b,t)}{v(\bar{b},t)}$$

which finishes the proof. The proof for the other direction of ratio monotonicity is similar. \textbf{Q.E.D.}

In other words, all products and bundles will have proportional demand curves and the exact same optimal quantity sold. As a result, the monopolist is indifferent among all possible bundling strategies.\footnote{Note that the above results assumed a ``deterministic'' version of ratio monotonicity. I add (without proving) that if we consider a ``stochastic'' version of ratio-monotonicity (a la \cite{haghpanah2019pure},) we can show that the two random variables $v(\bar{b},t)$ and $\frac{v(b,t)}{v(\bar{b},t)}$ would have to be independent of each other under weak ratio-monotonicity in either direction. Similarly, a strict form of ratio monotonicity would be infeasible. This implies that pure bundling is always optimal.}

Next, I turn to the implications of this paper's framework for additive values.

\begin{proposition}\label{prop: additive values thm 1}
Suppose values are additive and assumptions \ref{assumption: monotonicity}-\ref{assumption: quasiconcavity} hold. Then pure bundling is optimal if and only if $\forall b: D^*(b)=D^*(b^C)$.
\end{proposition}

\textbf{Proof.} If $\forall b: D^*(b)=D^*(b^C)$, then the monopolist will be indifferent among all possible bundling strategies. If $\exists b: D^*(b)>D^*(b^C)$, then one can use the fact that linearity implies  $D^*(b^C|b)=D^*(b^C)$, in order to show that $D^*(\bar{b})<D^*(b)$. Thus, pure bundling is not optimal. \textbf{Q.E.D.}

The fact that under assumptions \ref{assumption: monotonicity}-\ref{assumption: quasiconcavity} pure bundling is only optimal in such ``measure-zero'' cases should not be surprising. Additive values, once combined with monotonicity, will closely resemble s ``positive-correlation'' case in the \cite{adams1976commodity} context. We know that under positive correlation mixed bundling is preferable to pure bundling.

Overall, the tools originally developed to analyze bundling under non-additive values (including this paper) seem to be less powerful when restricted to environments with additive values.

\end{document}